# The photodetection of ultrashort optical pulse trains for low noise microwave signal generation

Franklyn Quinlan
National Institute of Standards and Technology, Boulder, CO 80305, USA
franklyn.quinlan@nist.gov

Abstract: Electrical signals derived from optical sources have achieved record-low levels of phase noise, and have demonstrated the highest frequency stability yet achieved in the microwave domain. Attaining such ultrastable phase and frequency performance requires high-fidelity optical-to-electrical conversion, typically performed via a high-speed photodiode. This paper reviews characteristics of the direct photodetection of optical pulses for the intent of generating high power, low phase noise microwave signals from optical sources. The two most popular types of photodiode detectors used for low noise microwave generation are discussed in terms of electrical pulse characteristics, achievable microwave power, and photodetector nonlinearities. Noise sources inherent to photodetection, such as shot noise, flicker noise, and photocarrier scattering are reviewed, and their impact on microwave phase fidelity is discussed. General guidelines for attaining the lowest noise possible from photodetection that balances power saturation, optical amplification, and amplitude-to-phase conversion, are also presented.

## 1. Introduction

The photodetection of ultrashort optical pulses is central to many RF and microwave photonics systems that wish to utilize the superb frequency and timing precision available from optical sources. The quality factor of optical resonators can exceed $10^{11}$ [1], yielding oscillators with fractional frequency instability below $5 \times 10^{-17}$ at 1 second and optical atomic clocks with fractional frequency stability at $1 \times 10^{-18}$ at a few thousand seconds [2, 3]. This is orders of magnitude better than the best microwave oscillators [4, 5]. Additionally, mode-locked laser sources can exhibit pulse-to-pulse timing jitter in the sub-femtosecond regime [6-8], as opposed to the 100s of femtoseconds to picosecond jitter of state-of-the-art microwave sources [9]. Transferring this optical stability to the RF, microwave, and millimeter-wave domain has applications in Doppler radar, whose sensitivity is enhanced with lower phase noise on the transmitted signal [10], in the use of microwave transmission for long-distance optical clock synchronization, in support of the future redefinition of the SI second with high-fidelity calibration of microwave clocks [11], and improving the sensitivity of very-long-baseline interferometry (VLBI), particularly when considering space-based VLBI systems [12].

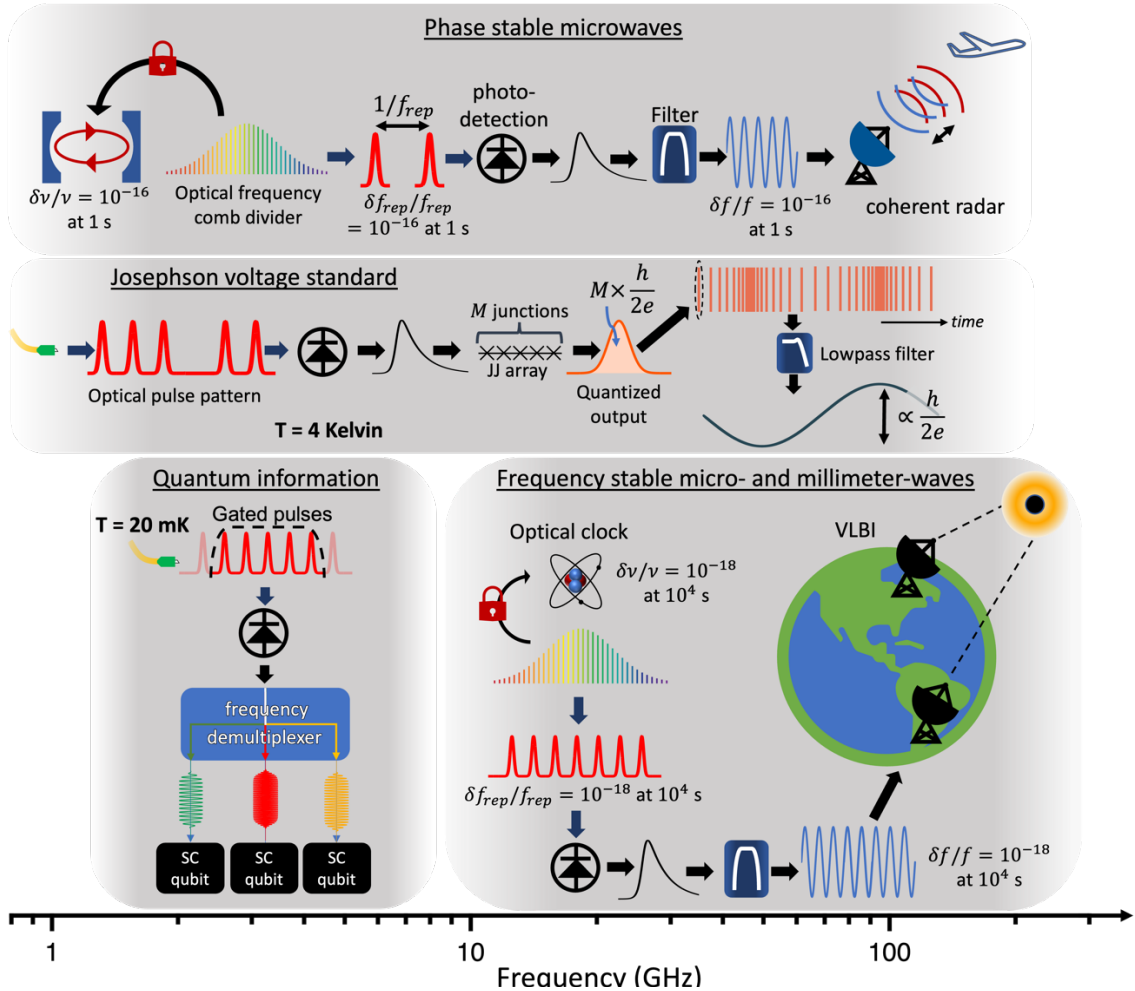

Fig. 1. Select applications of high-fidelity ultrashort optical pulse detection. The frequency scale at the bottom indicates the range over which the photodetector should operate for the given application. For example, quantum information processing with superconducting circuits (SC) tend to operate with microwave carriers below 10 GHz, whereas optically driven Josephson junction (JJ) arrays for quantum-based voltage standards may require electrical pulses with bandwidth from near-DC to 100s of GHz. VLBI; very long baseline interferometry.

There have been several demonstrations of the use of optical pulse detection to transfer the extraordinary level of timing and frequency stability of optical sources to the microwave domain. Examples include the generation of microwave signals that exhibit the lowest phase noise yet demonstrated in the microwave x-band [13-16], at 100 GHz [17], and at 5 MHz and 10 MHz [18]. Moreover, photodetection of pulse trains that are coherently linked to an optical atomic clock has yielded 10 GHz signals with fractional frequency instability at $1\times10^{-18}$ at $10^4$ seconds [19], two orders of magnitude more stable than the best microwave atomic clocks. There are additional applications that, while not requiring optical atomic clock-level stability, also take advantage of high-fidelity detection of ultrashort pulses. For example, photonic analog-to-digital conversion architectures use the lower timing jitter attainable from optical pulse trains to sample waveforms [20, 21], and short pulse detection can convey benefits in optical interconnects [22]. In coupled optoelectronic oscillators and regeneratively mode-locked lasers, short pulse detection is an integral part of the oscillator operation [23, 24]. High-fidelity optical pulse detection is also used as a diagnostic tool to measure optical pulse-to-pulse timing jitter [25, 26][27], the characterization of which is needed for a broad class of optical pulse-probe experiments. Further applications include optical two-way time transfer [28], and photonic links to cryogenic temperatures for applications in superconducting circuits [29], perhaps even aiding quantum information systems with superconducting qubits [30]. A few selected applications of low noise optical pulse detection are shown in Fig. 1.

Photodetection of ultrashort pulses to generate phase- and frequency-stable microwave signals places extremely stringent demands on the optical-to-electrical converter in terms of power handling and nonlinearity. Concerns over photodetection fidelity has led to the development of ways to extract the timing information of ultrashort pulses without direct detection of optical pulses [31, 32], where a microwave oscillator is phase locked to the optical pulse train through an electro-optic timing comparison. With this method, excellent results have been achieved in synchronizing an optical pulse train with a microwave oscillator [33], with residual noise comparable to the best optical-to-microwave conversion with direct photodetection [19, 34]. However, in direct detection there remains the possibility of broadband microwave generation, and allows for applications of the photocurrent pulses rather than a narrowband microwave signal.

In this paper we review developments in photodetector design and operation that have addressed both power saturation and noise in the pursuit of ever higher signal-to-noise ratio (SNR) in optically derived microwaves, providing orders of magnitude improvement in microwave phase noise in the past 10-15 years, and that have been an integral part in the production of microwave signals with phase and frequency stability that surpasses that of traditional microwave signal generators. Notably, both signal power and noise properties of short pulse detection differ from CW and modulated-CW illumination in important ways. We begin in the following section by considering the achievable microwave power, saturation characteristics, and nonlinearity of photodetectors under ultrashort optical pulse illumination, starting with a brief introduction of photodetector designs and circuits. Comparisons between microwave signal generation under short pulse detection and modulated-CW illumination are presented. Section 3 considers noise in photodetection, highlighting the properties unique to ultrashort pulse detection, and contrasting this behavior with CW illumination. This section also discusses photocurrent noise arising from optical amplification and the noise impacts of optical pulse interleavers. Section 4 provides guidelines for best operation for highest SNR in optically derived microwave signal generation, and in Section 5 we conclude with a look towards future developments and applications.

## 2. Photodetector Pulses, Microwave Power, and Nonlinearities
### 2.1 Photodiode designs and circuits
Photodetectors must contend with the often-competing requirements for high responsivity, high saturation power, high linearity, low intrinsic noise, and sufficient electrical bandwidth. For the photonic generation of microwave signals with large SNR, the most successful and widely available detectors are photodiodes. Several photodiode designs have been proposed and demonstrated to achieve the best performance in one or more desired metric, including pin detectors [35], dual-depletion region detectors (DDR) [36], uni-travelling carrier (UTC)-type detectors [37, 38], modified uni-travelling carrier (MUTC) detectors [39], and various waveguide detectors, including those based on pin, UTC and MUTC designs [40-42]. The vast majority of reported optically generated low noise microwaves have relied on DDR or MUTC detectors, and we therefore limit our discussion to these detector types. Detailed descriptions of DDR and MUTC detector designs can be found elsewhere [36, 39]; here we briefly introduce these two structures to help delineate their respective advantages for low noise microwave generation, as well as introduce aspects that help explain their response to ultrashort optical pulse illumination.

To understand the design choices in DDR and MUTC detectors, it is important to recall that the photodiode bandwidth is determined by two effects: the transit time of photoexcited electrons and holes, and the RC time constant given by the device capacitance and external circuit load impedance [35]. Band diagrams for the two detector types are shown in Fig. 2. The transit time is determined by the electron and hole mobilities and the length through which these carriers must travel from generation in the absorber to the diode n- and p-contacts. Dual-depletion region detectors may be considered as an extension of the pin detector design, where instead of the single depleted absorbing layer of the pin, the depleted region is separated into an absorbing and non-absorbing layer. Optical illumination produces electrons and holes in the absorber, which are swept towards the p-doped and n-doped regions due to the applied bias voltage. By placing the absorber near the p-contact, the slower holes only need to transit the absorber, while electrons transit both the absorber and nonabsorbing drift layer. Ideally the transit times for holes and electrons are balanced. The device capacitance may be understood as a simple parallel plate capacitor where a larger device diameter and thinner depletion length lead to higher capacitance. Thus adding a nonabsorbing drift layer does not reduce the transit-time limited bandwidth, but it does reduce the device capacitance. The end result is higher bandwidth and narrower impulse response. Compared to a pin photodiode, the saturation power is also increased, since it is not necessary to reduce the device diameter to achieve the desired capacitance for high-speed operation.

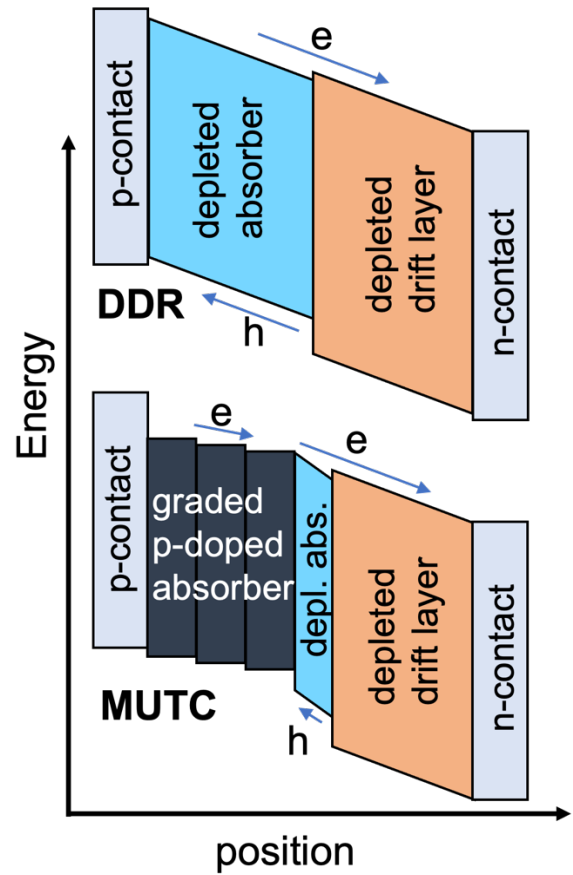

Fig. 2. Band diagrams of DDR (top) and MUTC (bottom) photodiodes. The DDR separates the depletion region into absorbing and non-absorbing layers in order to reduce the device capacitance without increasing the transit time of the holes. In the MUTC detector, most of the light absorption is in an undepleted p-doped layer, leading to a detector response dominated by electron transport dynamics.

MUTC detectors also employ separate depleted absorbing and nonabsorbing layers, with the additional inclusion of a p-doped, undepleted absorber. This further reduces the influence of hole transport, leaving the transport dynamics of the high mobility electrons as primarily responsible for the photodetector's response. To combat saturation effects, the MUTC additionally incorporates a multi-layered structure to tailor the electric field within the device. The electric field is shaped in a way that assists carrier transport as the photocurrent is increased, compensating for space-charge effects and helping the device maintain high-bandwidth operation [43, 44]. These innovations have led to the demonstration of 10 GHz power approaching 2 W under modulated-CW illumination for MUTC detectors [44]. Under short pulse illumination, 10 GHz power approaching 320 mW (25 dBm) has been demonstrated [45].

In addition to the device structure, the external photodiode circuit will also impact the available microwave power, photodiode bandwidth, and saturation characteristics. Two common equivalent circuits are shown in Fig. 3. In Fig. 3(a), an "internal" resistor, often 50 Ω, is placed near the photodiode [46], reducing the effective load resistance. This provides a reduction in the RC time constant and can increase the photodiode bandwidth. Moreover, the source impedance of the photodiode is transformed from the >10 MΩ of the photodiode shunt resistance to a level better matched to the external load. However, the available photocurrent is now divided across the internal resistor and the external load, reducing the available microwave power. Figure 3(b) shows an equivalent circuit without an internal resistor and where the bias voltage is introduced through a bias tee. Here, the output is necessarily AC coupled. For both circuits, we note that the maximum voltage swing across the load cannot exceed the photodiode bias voltage. This limit is observed in MUTC detectors under short pulse illumination with high peak power [47], and is discussed further in Section 2.2.1.

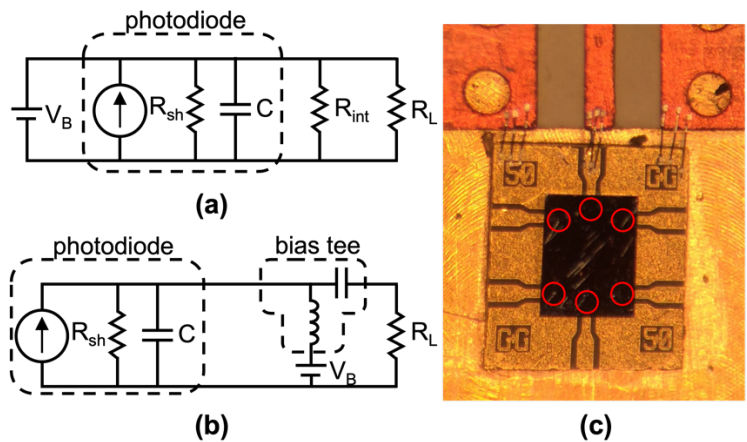

Fig. 3. (a) Photodiode equivalent circuit where an internal resistor, $R_{int}$, is included, often impedance-matched to the load $R_L$. (b) Equivalent circuit without an impedance-matching resistor and an AC coupled output. In both (a) and (b), the photodiode shunt resistance $R_{sh}$ is typically > 10 MΩ, and the photodiode capacitance C depends on the device diameter and thickness. An external bias is provided by source $V_B$. (c) Image of an array of six MUTC photodetectors that has been flip-chip bonded to a coplanar waveguide substrate with one detector wire-bonded to an external circuit. Red circles indicate individual photodetectors.

In a laboratory setting, connecting the photodiode structure to the external circuit can be achieved with microwave probes, while wire or flip-chip bonding to a co-planar waveguide is often used in packaged detectors (Fig. 3(c)). We note that in some cases inductive peaking can be optimized, leading to an increase the microwave power at chosen frequencies [48]. Stronger resonant enhancement can also be implemented at the desired microwave frequency for narrow band applications [49, 50].

2.2 Pulses, microwave power, and saturation

When considering ultrashort pulse illumination, a discussion of photodetector saturation must first differentiate whether the quantity of interest is the pulse peak voltage in the time domain or the power in a microwave harmonic in the frequency domain. While we are primarily interested in the frequency domain behavior when generating ultralow noise microwaves, examination of the time domain pulses offers important insights. We therefore begin our discussion of saturation by considering the time domain properties.

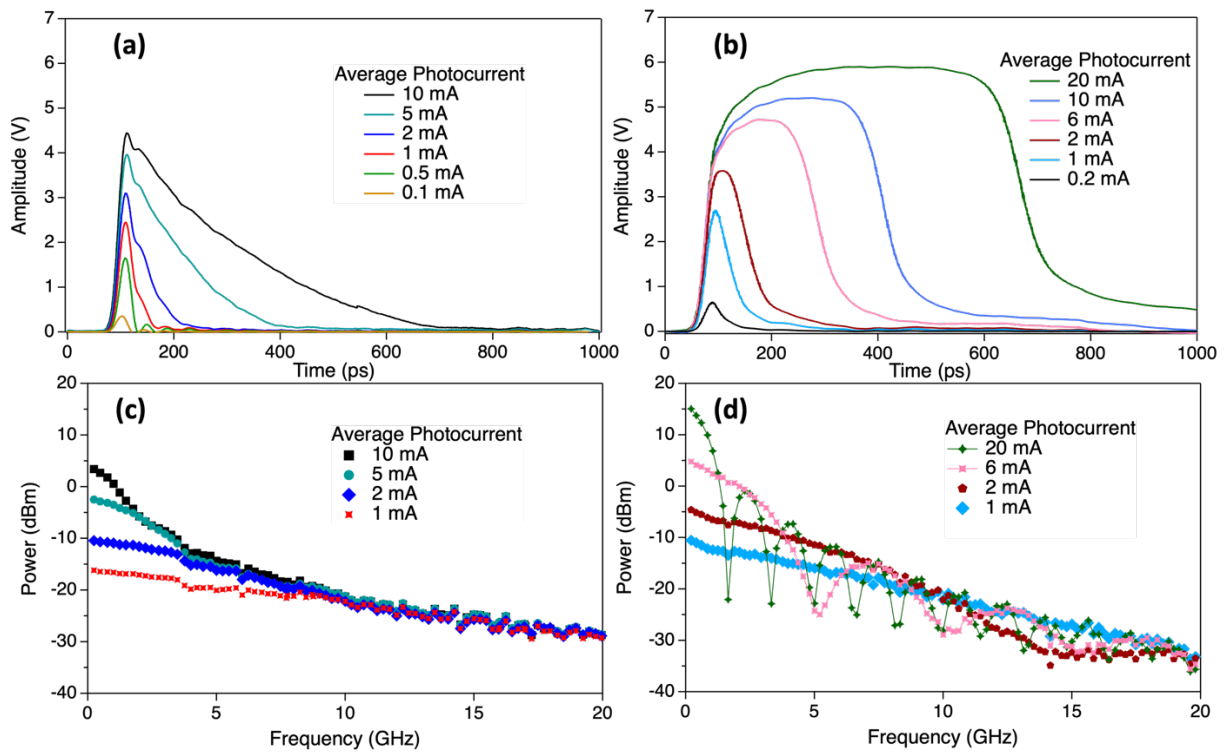

Fig. 4. DDR and MUTC photodetector response in time and frequency domains when illuminated by an ultrashort optical pulse train at 208 MHz repetition rate. (a) Sampling scope trace of the temporal response of the DDR detector. (b) Sampling scope trace of the MUTC temporal response. (c) DDR frequency response, and (d) MUTC frequency response, where every measurement point is a harmonic of $f_{rep}$. Powers given in (c) and (d) are those measured at the RF spectrum analyzer, and higher frequency powers may be reduced due to cable losses between the photodetector and the spectrum analyzer. Note that the DDR is packaged with an internal 50 Ω resistor. For (a) and (c) the DDR bias voltage is 8 V. For (b) and (d) the MUTC bias voltage is 7 V.

*2.2.1 Time domain*

Since the photocurrent is proportional to the optical power, detecting a train of optical pulses results in a train of electrical pulses. The electrical pulses are the convolution of the optical pulse with the detector's impulse response. The impulse response is governed by the transit time of the photoexcited carriers and the lumped-circuit RC time constant [35]. Importantly, the shape and width of the impulse response is strongly dependent on the energy of the incident optical pulse and the applied bias voltage. Full accounting of the nonlinear pulse distortion should consider transient space-charge and self-induced field effects, carrier bleaching, thermal effects, and the external load impedance over a broad bandwidth [51-53], but a few mechanisms tend to dominate. First, detecting a high energy optical pulse generates a large number of electron-hole pairs, inducing a space-charge field that counteracts the applied bias. This reduces the electric field strength in the depleted regions of the detector to a level below that required for the carriers to reach saturation velocity. This in turn increases carrier transit time and broadens the impulse response. Second, ultrashort optical pulses have a very large peak-to-average power ratio. For example, a 200 MHz pulse train of 1 ps duration pulses have a peak-to-average power ratio of ~5000. Without saturation, 10 mA of average photocurrent would imply a peak current of 50 A, or a peak voltage of 2.5 kV across a 50 Ω load. To avoid thermal damage and avalanche effects, the photodetector's applied bias voltage is kept well below this, clamping the peak voltage and distorting the pulse.

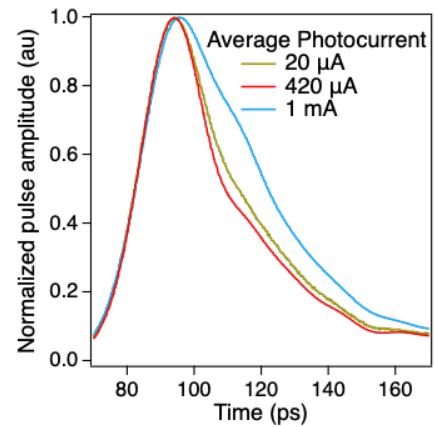

Fig. 5 MUTC normalized impulse response. As the photocurrent is increased, the full-width at half-maximum of the impulse response goes from 31 ps at 20 µA, to a minimum width of 27 ps at 420 µA, then begins to broaden, reaching 40 ps at 1 mA.

Details of the distortions of the impulse response strongly depend on the photodiode structure and loading of the external circuit, and are different for DDR and MUTC detectors. Figure 4 shows electrical pulses measured on a sampling scope for a DDR detector and an MUTC detector as a function of average photocurrent. Both detectors were illuminated by a train of sub-2 ps-wide optical pulses from an Er:fiber mode-locked laser operating at a repetition rate of 208 MHz. The average optical power was controlled by a variable attenuator before the photodetector. For both detectors there are significant changes to the pulse shape as the photocurrent increases. The DDR response is primarily due to space charge-effects that increase the transit time of the electrons and holes [54, 55], with a long response tail due to the lower mobility holes, as shown in Fig. 4(a). We also note that pulse width of a DDR broadens monotonically with average photocurrent [55].

MUTCs display a different temporal behavior, as shown in Fig. 4(b) and Fig. 5. Like the DDR, the creation of a space charge field influences the MUTC response; however, self-induced fields in the undepleted absorber and the predominance of electron transport dynamics also play important roles. The lack of hole transport is visible in Fig. 4(b), where the MUTC does not exhibit the same response tail as the DDR. Instead, under high-power illumination the MUTC's internal field collapses and the peak voltage at the load is limited to that of the applied bias voltage, resulting in a square-like pulse. Prior to pulse broadening, the pulse width passes through a minimum (seen more clearly in Fig. 5). This can be understood from carrier generation in the undepleted absorber creating a self-induced field that promotes carrier transport [56]. For the illumination used here, the pulse width is minimized at an average photocurrent near 420 µA, shown in Fig. 5. The electrical pulse width reduction, and associated bandwidth enhancement, is small, but has important consequences in the amplitude-to-phase conversion nonlinearity, discussed below in Section 2.4.

It is important to note that the average photocurrent for which the impulse response exhibits significant pulse distortion depends on the pulse repetition rate, $f_{rep}$, and the applied bias. That is, the impulse response distortion is fundamentally a function of the energy per pulse and the fields created within the detector relative to the field from the applied voltage bias. For example, peak voltages as high as 38.3 V for an MUTC detector with full-width at half-maximum of 30 ps at bias of 39.5 V has been demonstrated [57]. Illumination spot size also influences photodetector nonlinearities where tightly focused illumination increases the photocurrent density. Transforming a Gaussian spatial profile to a more uniform illumination [58], as well as defocusing to overfill the detector area (at the cost of lowered responsivity) have resulted in higher linearity [59].

*2.2.2 Frequency domain.*

In the frequency domain, a train of electrical pulses manifests as an array of discrete microwave tones whose frequencies are harmonics of $f_{rep}$. In the limit where the optical pulse width is much less than the impulse response of the detector, the microwave power, $P_\mu^{pulse}$, of each harmonic that is delivered to a load with impedance $R$ can be written in terms of the average photocurrent $I_{avg}$ as [26, 60]

$$P_\mu^{pulse} = 2I_{avg}^2 |H(f)|^2 R \qquad (1)$$

where $H(f)$ is the normalized Fourier transform of the detector's impulse response (such that $H(0) = 1$). As mentioned above, for detectors with internal impedance-matching resistors the photocurrent is split across internal and external loads. For equal internal and external impedances, the power across the load is reduced from that given by Eq. (1) by 6 dB.

The temporal broadening of the electrical pulses corresponds to a frequency-dependent saturation of the photonically generated microwave tones. In other words, $H(f)$ is power dependent, and tends to reduce in magnitude at high frequencies as the photocurrent increases. For the DDR detector response shown in Fig. 4(c), the power of the low frequency harmonics continues to increase with increasing photocurrent while the high frequency harmonics become clamped. The behavior of the MUTC detector, shown in Fig. 4(d), also displays a power increase at the low frequency harmonics. The overall frequency response, however, is quite different, where the square-like impulse response results in a sinc-like $H(f)$. Importantly, the peak voltage of the time domain pulse may saturate while low frequency harmonics of $f_{rep}$ continue to increase according to Eq. 1.

It is interesting to compare the microwave power under short pulse illumination to that given by modulated-CW illumination, expressed as

$$P_\mu^{CW} = \tfrac{1}{2}m^2 I_{avg}^2 |H(f)|^2 R \qquad (2)$$

where $m \leq 1$ is the modulation index. Thus photodetecting optical pulses can give 6 dB more microwave power than full depth of modulation on a CW optical signal with the same average photocurrent. Indeed, as long as the photodetected microwave harmonic is below the onset of saturation, short pulse detection gives the highest microwave power relative to the average (DC) power – given by $I_{avg}^2 R$ – of any waveform. Moreover, the fact that short pulse detection results in an array of microwave signals each with power given by Eq. 1 further enhances the delivered microwave power available from short pulse detection.

A closer examination of the microwave power for an MUTC detector, both with pulsed and modulated-CW illumination, is shown in Fig. 6. For short-pulse detection, the detector was illuminated by a mode-locked laser, and the power of the fundamental repetition rate at

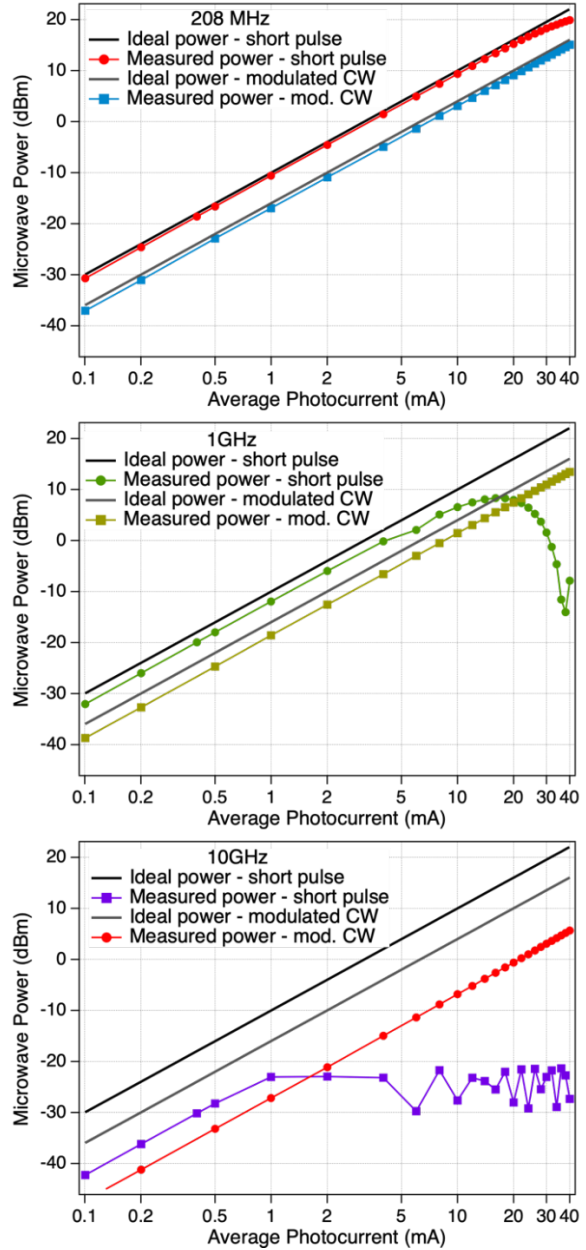

Fig. 6. The microwave power from photodetection of a train of ultrashort pulses with an MUTC detector at the pulse repetition rate fundamental (top), 5[th] harmonic (middle), and 48[th] harmonic (bottom). At each frequency, the measured power is compared to the ideal short-pulse microwave power (50 Ω load), the power generated by a CW laser that is modulated at the same frequency as the pulse train harmonic, and the ideal modulated-CW power. Small-signal deviations from the ideal powers at higher harmonics are due to the detector bandwidth roll-off as well as uncalibrated loss between the photodetector and RF power meter.

208 MHz, 5th harmonic near 1 GHz, and 48th harmonic near 10 GHz were recorded. Full depth of modulation ($m = 1$) on a CW signal was generated by beating two equal power, co-polarized lasers. The frequency separation between the lasers was set to match the measured microwave harmonics of the mode-locked laser, and the total optical power on the photodetector was controlled with a variable optical attenuator. The comparison of achievable microwave power at 208 MHz is shown in Fig. 6(a), where the microwave power for both short pulse and modulated-CW illumination hew close to the values predicted by Eq. 1 and Eq. 2, respectively. The power under short pulse illumination is ~6 dB higher for an average photocurrent below 20 mA. Above 20 mA, the microwave power for short pulses begins to saturate, but remains above the microwave power for modulated-CW up to photocurrents of at least 40 mA. Results at 1 GHz and 10 GHz are shown in Figs. 6(b) and 6(c), respectively. At these frequencies, cable loss and finite photodetector bandwidth reduce the measured powers, for both $P_\mu^{pulse}$ and $P_\mu^{CW}$, below the predicted ideal power levels. Importantly, the onset of microwave power saturation of $P_\mu^{pulse}$ occurs at much lower photocurrent than it does for the fundamental at 208 MHz, such that $P_\mu^{CW}$ can significantly exceed $P_\mu^{pulse}$. For example, measurements at 10 GHz in Fig. 6(c) shows $P_\mu^{CW}$ exceeding $P_\mu^{pulse}$ for an average photocurrent greater than ~ 2 mA, with a >30 dB advantage at 40 mA. More generally, the power advantage of short pulse detection is best realized for the lowest order harmonics while high order harmonics suffer from saturation even at moderate average optical powers.

2.3 Saturation relief with optical pulse interleaving

Microwave power saturation due to both space-charge effects and limited load voltage swing can be mitigated by increasing the photodetector's bias voltage. However, as mentioned above, this is limited by either avalanche breakdown, or runaway dark current as a result of excess localized heating [59]. Alternatively, microwave saturation power can be increased by increasing the pulse repetition rate to better match the desired microwave frequency (denoted $f_\mu$) [61]. A successful method to improve the microwave power from a low repetition rate laser is to multiply the optical pulse repetition rate through linear optical interferometry. This can be accomplished by different architectures [14, 61-63]. A common method, illustrated in Fig. 7(a), is to split the optical pulse train into two paths, place a relative delay of half the pulse repetition period between the two paths, and recombine. This results in a doubling of the pulse repetition rate, with the optical power split between the two output ports of the interferometer. The output ports can then be recombined with relative delay equal to ¼ of the pulse period to produce another doubling of $f_{rep}$. For $N$ cascaded interleavers, the resulting repetition rate is $2^N \times f_{rep}$. By halving the peak-to-average power ratio, the average photocurrent can double before the onset of saturation, leading to a microwave power increase of ~6 dB per stage, or 6*N* dB after *N* stages. Repetition-rate-doubling interleavers can be cascaded until the desired microwave frequency is an odd-harmonic of $f_{rep}$, at which point a final stage with a delay that is a multiple of $1/f_\mu$ can be added to provide a further boost to the microwave power [64, 65]. The 10 GHz power curves of Fig. 7(b) show the improvement in microwave power after five cascaded interleaver stages, where >25 dB increase in the saturation power is realized. For best power efficiency, the two outputs of the final interleaver stage can illuminate separate detectors and be combined electronically [66, 67], polarization multiplexing can be used in the final interleaver stage [14, 68], or ring interferometer designs can be implemented [62].

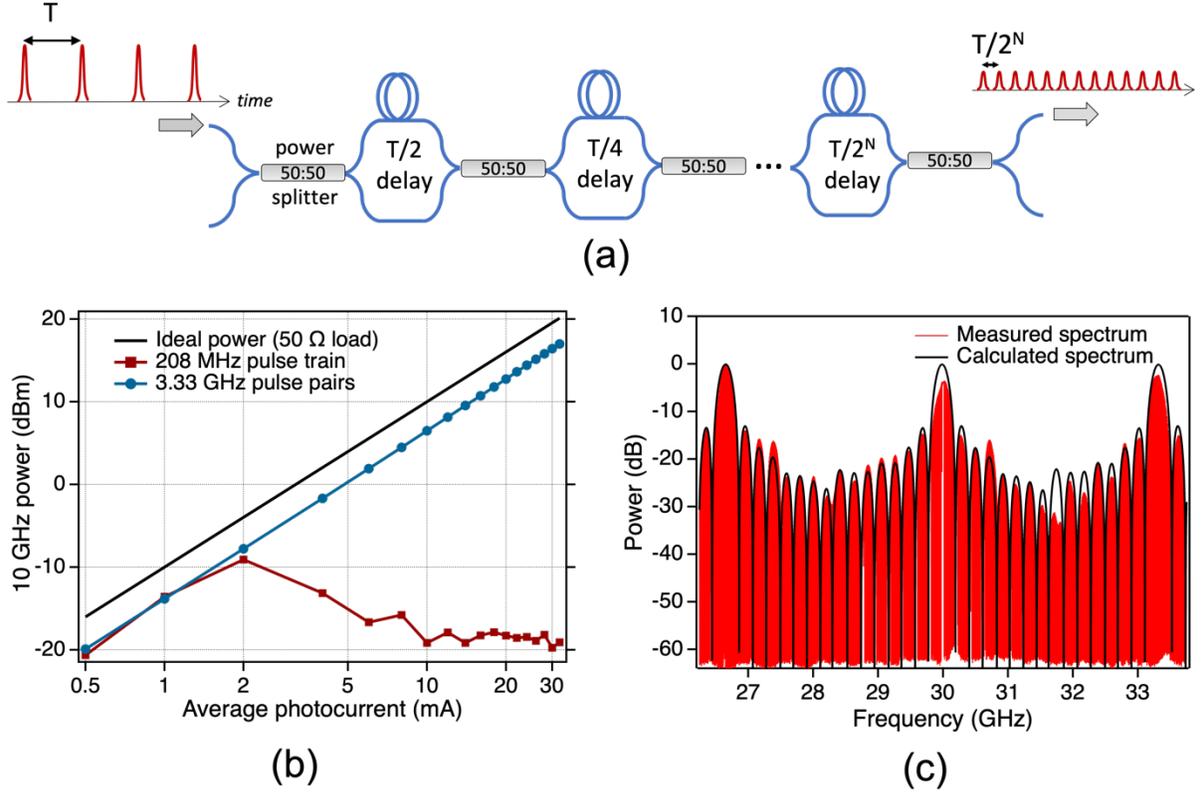

Fig. 7. (a) Optical pulse interleaver schematic. A pulse period of T at the input is reduced to T/2^N for N stages when each stage halves the pulse period. (b) Power in the 10 GHz harmonic from a 208 MHz pulse repetition rate with (blue circles) and without (red sqaures) an optical pulse interleaver. A 3.33 GHz pulse train is produced after four interleaver stages. The fifth and final interleaver stage uses a 100 ps delay, producing pairs of pulses at a repetition rate of 3.33 GHz. Ideal power is into a 50 $\Omega$ load. (c) Spectrum of the 5-stage interleaver transfer function, measured by scanning the pulse repetition rate and using a spectrum analyzer set to maximum hold. The calculated spectrum is shown for comparison.

The full transfer function of an optical pulse interleaver (defined as the spectrum of the output optical power profile divided by the spectrum of the input optical power profile) is easily derived when the interleaver is recognized as a delay line filter [69]. Each stage simply duplicates the input pulse with delay $t_d$. In general, there may be amplitude imbalances between the pulses traveling through separate interleaver arms, and dispersion in different path lengths may lead to unequal optical pulse widths. Assuming Gaussian pulses at the interleaver input, the transfer function of a single interleaver stage that accounts for unequal pulse amplitude and pulse widths is

$$G(f) = \alpha e^{\pi^2 f^2 [\tau_1^2 - \tau_{in}^2]} + \beta e^{\pi^2 f^2 [\tau_2^2 - \tau_{in}^2]} e^{i2\pi f t_d} \qquad (3)$$

Where $\tau_{in}$ is the input pulse width, $\tau_1$ and $\tau_2$ are the widths of the pulses after traveling separate dispersive paths, and $\alpha$ and $\beta$ are coefficients to account for unequal pulse powers. In many cases, however, the dispersion is small enough that pulse broadening may be neglected, and the

power is well balanced between arms of the interleaver. In this case the transfer function reduces to

$$G(f) = 1 + e^{i2\pi f t_d} \tag{4}$$

This transfer function can be multiplied by an overall loss factor, ideally equal to 1/2 for an interleaver with two outputs. The transfer function for *N* cascaded interleaver stages is then

$$G_{TOT}(f) = G_N \cdots G_1 \tag{5}$$

where again an overall factor of 1/2 can be applied when the interleaver has two outputs. Upon photodetection, the microwave power spectrum is modulated by $|G_{TOT}(f)|^2$. This transfer function produces a series of peaks and nulls whose positions are determined by $t_d$. A single-stage interleaver will exhibit a sinusoidal frequency response with null frequencies of

$$f_m = \frac{2m-1}{2t_d} \tag{6}$$

For positive integer *m*. When $1/t_d$ is half the input pulse repetition rate $f_{rep}$, the interleaver produces a null at the odd harmonics of $f_{rep}$, resulting in harmonics spaced by $2f_{rep}$ as expected. When properly normalized the peak value of $G$ is unity, such that the power in a microwave harmonic is no more than that given in Eq. 1. (The value of the interleaver is that it extends the photocurrent range over which Eq. 1 is valid.) Subsequent interleaver stages will add nulls at frequencies corresponding to the individual stage delay. Figure 7(c) shows part of the spectrum of a photodetected pulse train after a 5-stage pulse interleaver. The calculated spectrum is also shown, showing good agreement with the experiment, despite neglecting in the calculation the pulse width imbalances from the interleaver's dispersive fiber couplers.

An error in the delay $t_d$ of an interleaver will shift the peak of the interleaver transfer function, introducing a misalignment with the microwave frequency of interest. This reduces the microwave power and consequently the microwave SNR at the thermal noise limit. However, as discussed in more detail in Section 3.1c, the impact to the shot noise-limited phase noise is much more severe and can place a tighter constraint on interleaver timing errors than the thermal noise.

2.4 Amplitude-to-phase conversion
In addition to saturation, high-speed photodetectors can display other nonlinearities, often characterized in terms of harmonic distortion, intermodulation distortion and amplitude-to-phase (AM-to-PM) conversion [51, 70]. In low noise microwave generation via the detection of ultrashort optical pulses, the most relevant characterization of nonlinearity (in addition to saturation) is AM-to-PM conversion, where optical pulse-to-pulse energy fluctuations are converted to microwave phase fluctuations. Indeed, photonically generated microwaves can exhibit phase noise that is several orders of magnitude lower than the intensity noise [15, 65]. Thus any transfer of optical intensity noise to microwave phase noise through AM-to-PM conversion can severely degrade the microwave phase noise performance. The importance of AM-to-PM conversion in photodetection has led to considerable attention towards understanding its properties and mitigating its impact on low noise microwaves [46, 52, 53, 55, 66, 71-77].

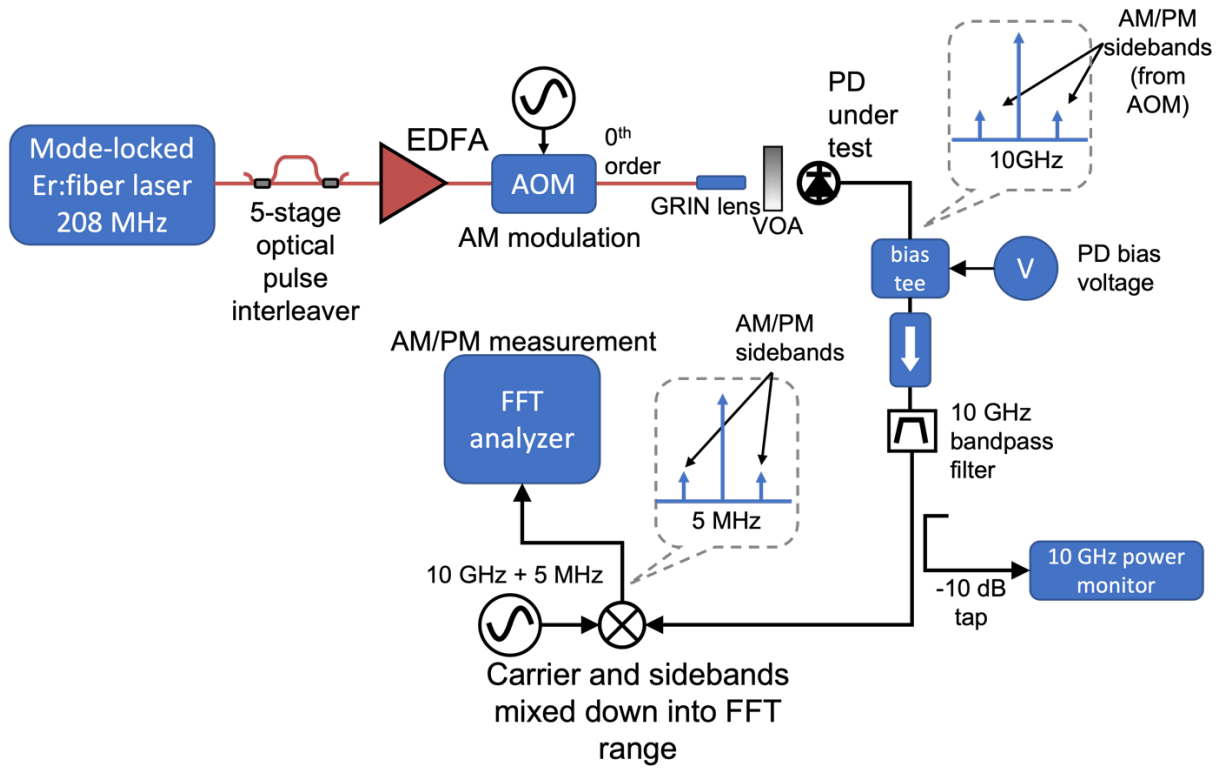

Fig. 8. Amplitude-to-phase conversion measurement setup. The optical power of a pulse train is modulated at a low rate prior to illuminating a photodetector under test. The optical power modulation results in amplitude and phase modulation of the photonically generated microwaves. One of the microwaves is selected for measurement of the relative level of phase modulation to amplitude modulation. AOM, acousto-optic modulator; FFT, fast Fourier transform; PD, photodetector; EDFA, erbium-doped fiber amplifier; VOA, variable optical attenuator.

The same physical mechanisms of space-charge and self-induced fields that produce saturation and pulse distortion are largely responsible for AM-to-PM conversion in DRR and MUTC photodiodes. Since the strength of these fields varies with optical power, the speed of the photocarriers also varies with optical power, coupling power and timing. A sign of reduced photocarrier velocity is the broadening of the impulse response and accompanying saturation of microwave power. Thus large AM-to-PM conversion would be expected for a photodetector operating in saturation. However, significant levels of AM-to-PM conversion can also happen at photocurrents well below saturation [52, 55, 72, 73, 75, 76]. Moreover, photodetectors operating at saturation photocurrent levels can exhibit a sign change in the AM-to-PM conversion yielding occasional nulls in the AM-to-PM magnitude, and an AM-to-PM null will appear below saturation for MUTC detectors. In this section, we discuss how such AM-to-PM conversion behavior is measured, characterized, and interpreted in DDR and MUTC detectors.

A number of techniques have been used to quantify AM-to-PM conversion in PDs [46, 52, 55, 72-74, 78]. Figure 8 illustrates one technique where the optical power is weakly modulated at a rate of $f_{mod}$ in the range of 10 kHz to 100 kHz, followed by a variable optical attenuator (VOA) to set the average optical power illuminating the PD under test. The phase modulation at $f_{mod}$ is then measured on a harmonic of the pulse repetition rate. The AM-to-PM conversion is characterized as the resulting root-mean-square phase shift due to a fractional change in the optical power [73]. The AM-to-PM conversion can then be characterized as a function of the average optical power illuminating the photodetector by tuning the VOA. Measuring the AM-to-PM conversion in this way allows us to directly calculate the influence of the laser's relative intensity noise (RIN) on the microwave phase noise, as

$$L^{RIN}(f) = S^{RIN}(f) + 20 log(\alpha) - 3 \qquad (7)$$

where $S^{RIN}(f)$ is the laser RIN power spectrum in units of dB/Hz, $\alpha$ is the AM-to-PM conversion coefficient that gives the measured rms phase deviation in relation to the fractional amplitude change imparted optically, given in units of $rad_{rms}/(dP/P)$, and $L^{RIN}(f)$ is the single-sideband phase noise due to RIN. As a low-RIN mode-locked laser can have a RIN value around -140 dB/Hz at 10 kHz offset [15, 65], an AM-to-PM coefficient < -40 dB is typically desired for low noise microwave generation.

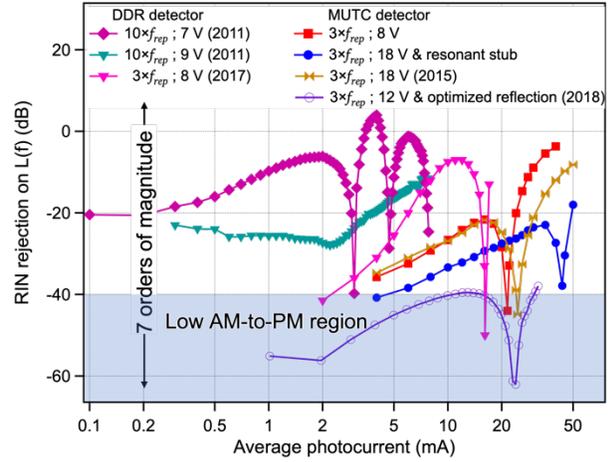

Fig. 9. Survey of AM-to-PM conversion for DDR [73, 74]. and MUTC detectors [34, 53].

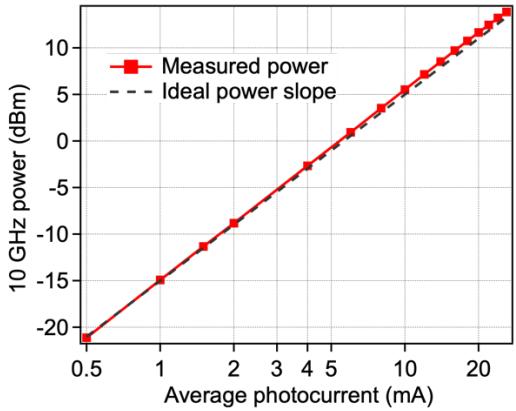
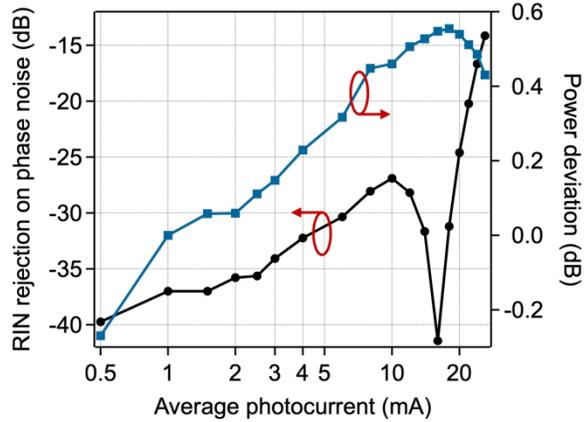

(a) (b)

Fig. 10. (a) Measured 10 GHz power for an MUTC detector as a function of photocurrent, showing deviation from $I_{avg}^2$ dependence. At moderate photocurrent, the detector bandwidth increases with photocurrent and the microwave power increases at a rate faster than predicted by a simple $I_{avg}^2$ dependence. (b) Comparison between AM-to-PM nonlinearity and microwave power deviation, where the maximum deviation is observed to align with the AM-to-PM conversion null. Power deviation level value arbitrarily shifted to zero at 1 mA average photocurrent.

Representative AM-to-PM values for DDR and MUTC photodetectors are shown in Fig. 9. Common to all the detectors in Fig. 9 is the existence of a photocurrent level where the AM-to-PM coefficient drops dramatically – the so-called AM-to-PM null. Device models, empirical photocurrent pulse models, and measurements of the AM-to-PM phase have shown these null points in the AM-to-PM magnitude are where the sign of the AM-to-PM coefficient changes [52, 53, 66, 74, 75]. Provided the operating photocurrent is at near the null point, AM-to-PM conversion below -40 dB is possible with any of the detectors shown in Fig. 10. Given Eq (1), we desire low AM-to-PM at as high an average photocurrent as possible so as to produce as high-power microwave signal as possible with phase noise uncorrupted by RIN. Also, the photocurrent range that gives an acceptably low AM-to-PM value can be extremely narrow for some detectors, making it difficult to maintain AM rejection over extended periods of time. Balancing the AM-to-PM from two detectors has shown to be an effective means to increase the photocurrent range [66], and feedback control can be implemented to maintain operation at the null point [74]. Alternatively, photodetector designs with higher linearity relax the requirements on maintaining a specific photocurrent level [52, 53]. The AM-to-PM of the highest linearity MUTC device shown in Fig. 9 remains below 40 dB for any photocurrent up to 30 mA. This implies microwave powers can approach 20 dBm while maintaining low AM-to-PM nonlinearity.

There is an important distinction in the AM-to-PM behavior in DDR photodetectors and MUTC photodetectors. A DDR photodetector will exhibit its lowest photocurrent AM-to-PM null at a photocurrent beyond which the microwave power saturates, and the electrical pulse width is significantly broadened [55, 73, 74]. In contrast, the photocurrent at which an MUTC detector displays its first null is below that where the microwave power saturates and can be linked to its bandwidth enhancement. As mentioned above, the electric field in MUTC detectors is tailored to increase the saturation power in such a way that the device bandwidth is enhanced at moderate photocurrents. At the photocurrent level where the bandwidth is maximized, the photocarrier speed is also maximized. At higher photocurrents, photocarriers begin to slow due to the large space-charge field with a commensurate reduction in the device bandwidth. At this "turn-around" photocurrent, where the photocarrier speed changes from increasing to decreasing, the AM-to-PM conversion is expected to pass through zero. This connection between bandwidth enhancement and AM-to-PM nonlinearity is shown in Figure 10. An MUTC photodetector was illuminated by a 3.33 GHz pulse train. The maximum bandwidth is determined by close monitoring of the microwave power of the 3$^{rd}$ harmonic at 10 GHz as a function of average photocurrent. Since the microwave power is proportional to the square of the photocurrent, a log-scale plot of the microwave power is expected to increase linearly with photocurrent with a slope of 2. With a bandwidth enhancement, the slope will exceed 2 as the photocurrent increases. Fig. 10(a) shows the measured microwave power with a slope 2 line also shown for comparison. Though small, there is a clear deviation in the measured power slope. Figure 10(b) shows the measured power deviation from a slope-of-2 increase. The power deviation peaks at the photocurrent of maximum bandwidth, and the AM-to-PM conversion is expected to exhibit a minimum at this same photocurrent. The photocurrent of the AM-to-PM conversion minimum is indeed measured to be close to the photocurrent at the maximum power deviation, as also shown in Fig. 10(b). These data indicate power measurements can be a convenient proxy for AM-to-PM measurements in MUTC detectors. Furthermore, it is important to note that the AM-to-

PM conversion is smallest at the narrowest electrical pulse width. As discussed below, other phase noise sources are also minimized for narrow pulses, producing a fortuitous low-noise combination in MUTC photodetectors.

Just as higher bias voltage and more uniform illumination tend to reduce microwave power saturation, they also reduce AM-to-PM conversion in both DDR and MUTC detectors. Additionally, we note that the broadband impedance of the external circuit will impact the AM-to-PM conversion, particularly at the null photocurrent, as first pointed out in [72]. Microwave components such as filters and isolators can reflect out-of-band frequency components back toward the photodiode. Upon reflection, out of band $f_{rep}$ harmonics will mix in the photodiode, generating new frequency components. The harmonic relationship of the microwave tones ensures that any nonlinear mixing product will lie at the frequency of some other harmonic of $f_{rep}$, producing a phase shift of that harmonic. This property was explored in detail in [53], where the phase of the strongly reflected $f_{rep}$ fundamental at 3.33 GHz and 2$^{nd}$ harmonic at 6.67 GHz were used to shift the AM-to-PM of the third harmonic at 10 GHz by as much as 30 dB. In some cases, the reflected harmonics were controlled to remove the existence on an AM-to-PM null altogether.

## 3. Photocurrent Noise

In this section, we describe measurements and analysis of the noise sources inherent to microwave signals generated via the photodetection of ultrashort optical pulses. Noise in photodetection may be separated into close-to-carrier flicker noise, and white noise from shot noise, thermal noise, and photocarrier transport. (Excess technical noise due to, say, a noisy bias voltage supply, may also impact the resulting microwave noise, but is not considered here.) Key points that distinguish the behavior of the photocurrent noise in short pulse detection are (1) in the detection of ultrashort optical pulses, shot and flicker noise do not equally contribute to the phase and amplitude noise of the generated microwave, rather the noise primarily lies in the amplitude quadrature; (2) for high power handling photodetectors, optical amplification may be beneficial, but there are optical pulse width-dependent impacts on the microwave phase noise from the optical amplifier as well; and (3) errors in the pulse delay from optical pulse interleavers can increase the shot and optical amplifier noise-limited phase noise floor. As described below, these pulse timing-dependent behaviors can be understood as a consequence of the time-varying (nonstationary) noise statistics inherent in short pulse detection.

Photodetection noise terms that impact the phase of a microwave signal derived from an optical pulse train are summarized in Fig. 11. This section begins with descriptions of the noise at the shortest timescales, that is, the far-from-carrier white noise floor, before turning to flicker noise. We note the analysis of noise under short pulse illumination is generally applicable to any device structure; the noise only depends on the device details insofar as the device determines the electrical pulse width and saturation power.

## 3.1 White noise

### 3.1.1 Thermal noise

With exception of extremely cold temperatures where the quantum nature of the microwave field is manifest [79], an electrical circuit at temperature $T$ will exhibit broadband thermal, or Johnson, noise at a level of $k_b T$ (W/Hz), where $k_b$ is Boltzmann's constant. The presence of this noise restricts the achievable SNR of a microwave signal. Thermal noise modulates the amplitude and phase noise of the generated microwave, contributing equally to both quadratures. The thermal noise contribution to the single-sideband microwave phase noise in 1 Hz bandwidth is expressed as [80]

$$L_{thermal} = k_b T / 2P_\mu. \qquad (8)$$

From Eq. (8) it is apparent that the more microwave power available from the photodetector, the lower the thermal noise-limited phase noise. For a photodetector that is not in saturation, the 6 dB microwave power improvement of short pulse detection over modulated-CW detection improves the thermal noise-limited SNR an equal amount. With their ability to greatly increase the microwave saturation power, optical pulse interleavers clearly can be an important tool to increase the SNR.

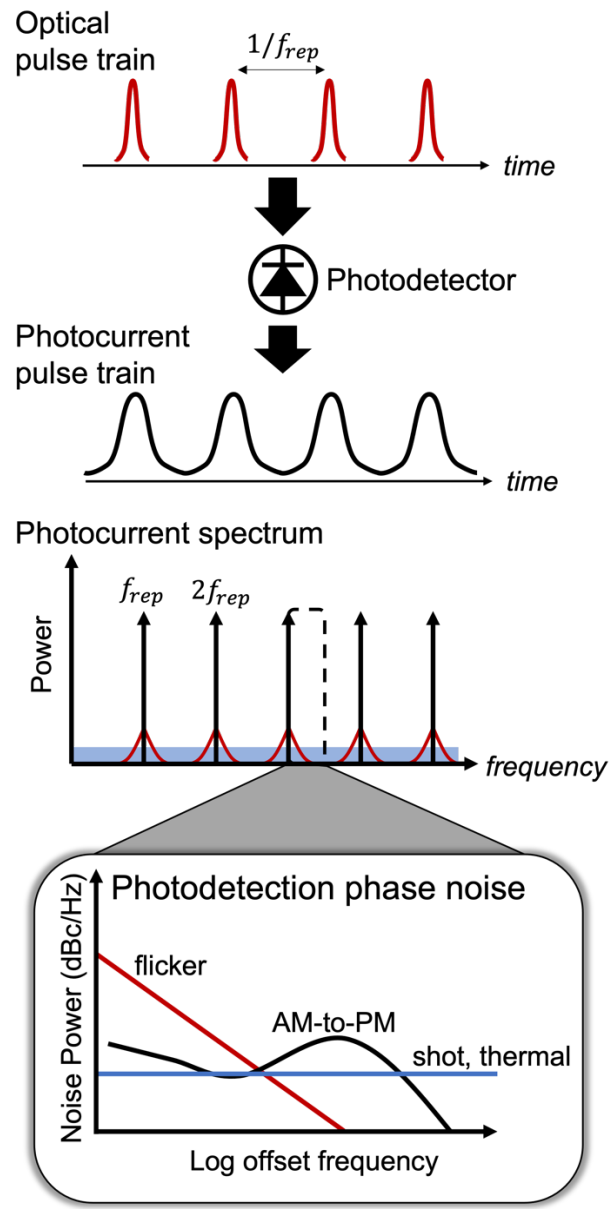

Fig. 11. Photodetection of a train of ultrashort optical pulses results in a train of electrical pulses. In the frequency domain, the train of electrical pulses is manifest as an array of equally spaced tones at the pulse repetition rate and its harmonics. While photodetection can transfer optical stability to the microwave domain with extremely high fidelity, noise is added in the photodetection process in the form of flicker noise, amplitude-to-phase noise conversion, as well as white noise such as shot noise and thermal noise.

### 3.1.2 Shot noise and photocarrier scattering

Like thermal noise, shot noise is broadband, and will modulate the amplitude and phase of a photonically generated microwave. Unlike thermal noise, photocurrent shot noise is not in general equally divided between the microwave signal's amplitude and phase noise [81]. To describe the shot noise behavior, we consider a simplified model of photodetection as shown in Fig. 12 [35, 82]. Illumination of the detector generates photocarriers with probability proportional to the incident optical power. These photocarriers then transit the detector, creating current impulses in the external circuit. The sum of these current impulses is the total photocurrent. To understand the noise, we divide the generation of photocurrent into two distinct processes: the generation of photocarriers and the subsequent photocarrier transport. In what follows, we reserve the term "shot noise" to refer to the stochastic process of photocarrier generation, and the term "photocarrier transport noise" to refer to all other noise effects related to photocarrier transport, such as carrier scattering events.

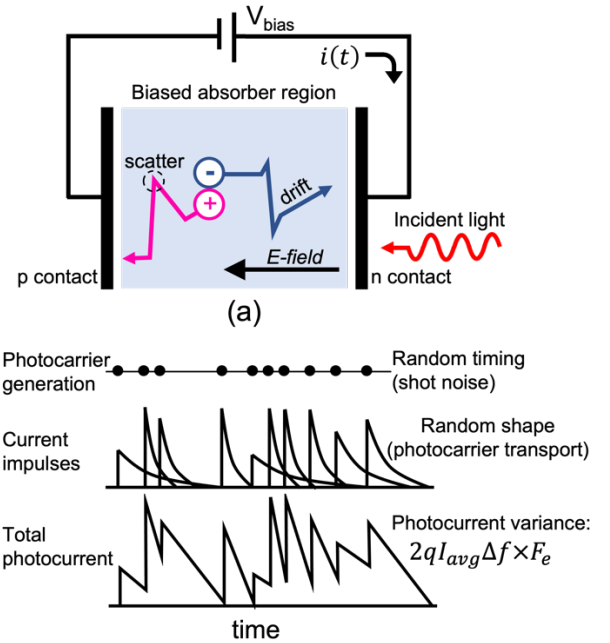

Fig. 12. Photocurrent model for shot and photocarrier scattering noise.

First, we consider the shot noise resulting from the Poissonian nature of photocarrier generation. In the absence of any squeezing of the optical field, the shot noise power in 1 Hz bandwidth around frequency $f$ may be written as [83]

$$P_{shot} = 2qI_{avg}|H(f)|^2 R \qquad (9)$$

where $q$ is the electron charge. This expression holds for CW, modulated-CW, or short pulse detection. When generating a microwave signal via the detection of a time-varying optical signal, this noise modulates the amplitude and phase of the microwave. Importantly, the shot noise distribution between the microwave amplitude and phase quadratures depends on the optical duty cycle [60, 81]. In the detection of ultrashort optical pulses, the shot noise contribution to the microwave phase can be orders of magnitude lower than its contribution to the amplitude noise. This can be viewed as a result of the cyclostationary nature of the shot noise for a train of ultrashort pulses, where the noise statistics are periodic with the pulse repetition rate [84]. A semi-classical time domain interpretation is given in Fig. 13(a), where we recall that frequency domain phase noise is manifest as timing jitter. The Poissonian statistics of light detection implies the variance in the number of generated photocarriers is proportional to the mean photocarrier number [35]. When detecting a train of ultrashort pulses, there will be a pulse-to-pulse variation in the number of generated photocarriers, leading to a time-averaged photocurrent noise variance of $2qI_{avg}\Delta f$ for measurement bandwidth $\Delta f$. (The equivalent noise power spectral density is given by Eq. (9).) Importantly, the temporal distribution of photocarriers within a pulse will also vary, resulting in slight shifts in the pulse center, or, in other words, timing jitter. The magnitude of the temporal shift is necessarily smaller for short pulses than for long pulses. We can also view the timing jitter in terms of a measurement, where we compare (multiply) the zero-crossings of a timing reference to the pulses' arrival. For shorter optical pulses, the noise is compressed to instances very near the zero-crossing of the reference. Multiplying with the reference signal results in a lower time-averaged product, or lower timing noise. Note that this description is equivalent to finding the pulse "center of mass" [85], where the time of arrival of a pulse $t_p$ can be defined as

$$t_p = \frac{\int t \times P_{opt}(t)\, dt}{\int P_{opt}(t)\, dt} \qquad (10)$$

Here $P_{opt}$ represents the optical pulse power profile. Multiplying the pulse by the linear slope of the timing reference near its zero-crossings acts as the $t$ multiplier in Eq. (10). Repeating this over multiple pulses determines the jitter.

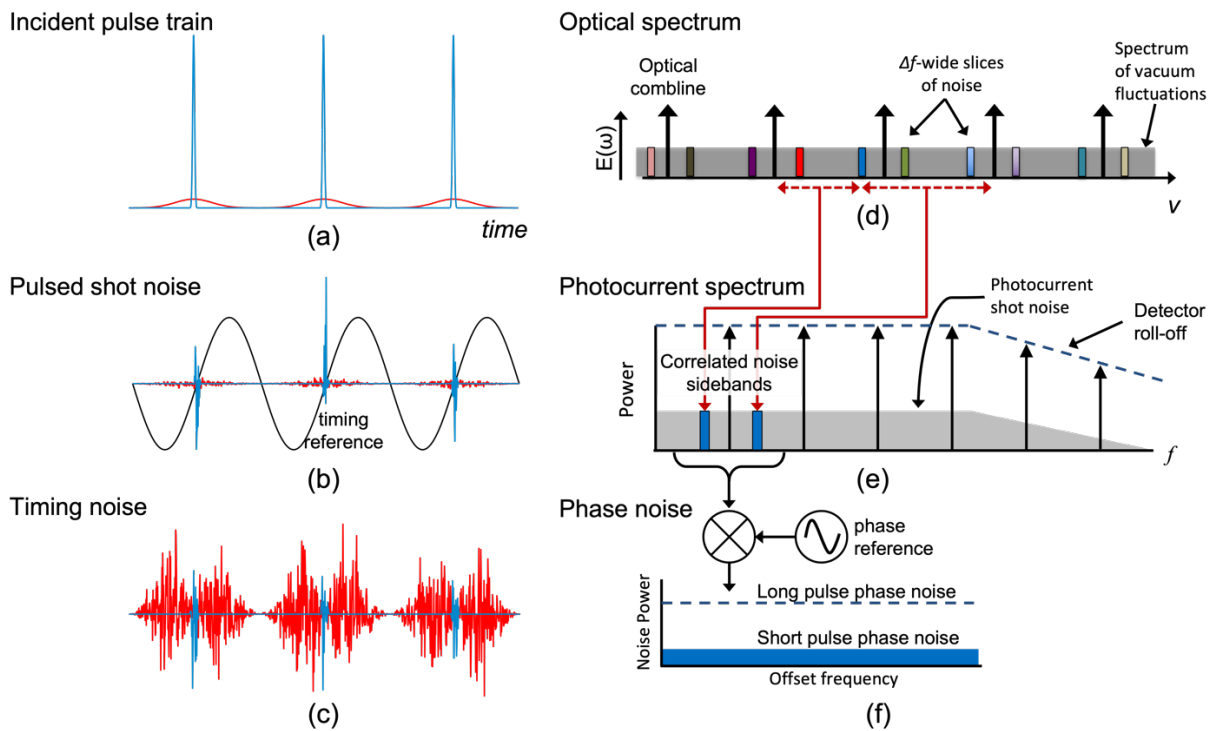

Fig. 13. Shot noise from pulsed detection and its contribution to timing and phase noise. (a) Two incident pulse trains with equal energy per pulse and average power, but with different pulse widths. (b) Time distribution of the shot noise of the pulse trains in (a), along with a timing reference. (c) Resulting timing noise of short and long pulses, obtained by multiplying the shot noise with the timing reference signal. The time-averaged timing noise is much lower for short pulses. (d) Optical spectrum of a train of pulses, with comb lines separated by the pulse repetition rate and where the shot noise is envisioned as a white noise spectrum populated at the vacuum level of ½ photon per mode. (e) Heterodyne beating between the vacuum modes and optical comb lines results in photocurrent shot noise. The correlated phase among the optical modes results in correlated photocurrent shot noise sidebands symmetric about a harmonic of the pulse repetition rate. (f) Phase noise measurement, where the higher degree of noise correlation for short pulses results in lower phase noise.

A frequency domain description of shot noise utilizing spectral correlations also provides insight [81]. This interpretation starts with an examination of the optical spectrum corresponding to a train of optical pulses, shown in Fig. 13(d), represented as a frequency comb. Importantly, the phase relationship among the comb lines is well-defined, and for transform-limited pulses all comb lines are in phase. Vacuum fluctuations are represented as a continuous background, populated by ½ photon per mode [86, 87]. In this picture, heterodyne beating between comb lines results in signals at harmonics of $f_{rep}$, and the beating between the comb and vacuum fluctuations results in photocurrent shot noise. If we consider a narrowband slice of noise in the optical domain that is within the comb bandwidth, this noise slice is translated to multiple photocurrent frequencies via heterodyning with multiple comb lines. Since the comb lines are phase correlated, the noise at different photocurrent frequencies is also correlated, including frequencies symmetric about the harmonics of $f_{rep}$ (Fig. 13(e)). The level at which the noise sidebands contribute to the amplitude or phase noise of the carrier depend on their relative phase, with in-phase sidebands producing amplitude modulation, and out-of-phase sidebands producing phase modulation (uncorrelated sidebands contribute equally to amplitude and phase). The fact that the optical comb lines are in-phase implies the photocurrent shot noise sidebands will also be in-phase, resulting in amplitude noise.

For optical pulses of finite width, some of the noise will end up in the phase quadrature due to two effects. First, the bandwidth of the comb is limited, and therefore part of the noise at any particular photocurrent frequency will be the result of a vacuum mode that resides outside the comb bandwidth and only beats with a single comb line. This produces partially uncorrelated sidebands about the microwave carrier, contributing equally to the amplitude and phase. The fraction of the uncorrelated noise depends on the comb bandwidth, such that broader bandwidths exhibit a higher degree of microwave sideband correlation. Second, the optical pulses may not be transform-limited, modifying the relative phase among comb lines and thereby shifting the relative phase between sidebands away for pure amplitude modulation.

An analytical expression for the shot noise contribution to the amplitude and phase noise of a photonically generated microwave has been derived [60]. For Gaussian shaped optical pulses, the phase noise can be expressed as

$$L_{shot} = \frac{qI_{avg}|H(f)|^2 R}{P_\mu}\left[1 - \exp\left\{-\left(2\pi f_\mu \tau_{opt}\right)^2\right\}\right] \quad (11)$$

where $\tau_{opt}$ is the $1/e$ half-width of the optical pulse intensity profile (a function of both the optical bandwidth and pulse chirp). The first part of the noise expression is the ratio of the single sideband shot noise to the microwave power. The second part of the expression in brackets gives the pulse width dependence, the exact form of which depends on the pulse shape. With Gaussian pulses, we may further express $P_\mu$ to account for a decrease in microwave power with broader optical pulses as

$$P_\mu = 2I_{avg}^2|H(f)|^2 R \times \exp\left\{-2\left(\pi f_\mu \tau_{opt}\right)^2\right\} \quad (12)$$

Substituting Eq. (12) into Eq. (11) gives the full expression. In the limit of short optical pulses where $f_\mu \tau_{opt} \lesssim 0.1$, the shot noise-limited phase noise may be simplified to

$$L_{shot}^{pulse} = \frac{q}{2I_{avg}} \left(2\pi f_\mu \tau_{opt}\right)^2 \qquad (13)$$

where the strong reduction of phase noise for short optical pulses is apparent. As we have defined shot noise as only determined by photocarrier generation statistics, it is the optical pulse width, not the much longer impulse response time of the photodetector, that determines the phase noise level at the shot noise limit.

For comparison, the shot noise-limited phase noise for a microwave signal generated with modulated-CW light is given by

$$L_{shot}^{CW} = \frac{2q}{I_{avg}} \qquad (14)$$

Where Eq. 2 is used along with Eq. 9, assuming 100% depth of modulation. In this case the shot noise is equally divided between amplitude and phase quadratures. Compared to short pulse illumination, the phase noise is further increased by the 6 dB lower microwave power for a given average photocurrent.

The phase noise implied by Eq. 13 is quite low: for a train of 1 ps Gaussian-shaped optical pulses generating 10 mA of average photocurrent, the shot noise-limited phase noise is below -200 dBc/Hz. The thermal noise at the same 10 mA photocurrent is -187 dBc/Hz at room temperature. This represents a significant improvement over the expected phase noise floor from a sinusoidally modulated CW laser generating the same photocurrent, which according to Eq. 14 is shot noise limited at -165 dBc/Hz. With the possibility of more than 20 dB increase in the dynamic range of the phase noise compared to modulated-CW illumination, it becomes important to consider the stochastic nature of photocarrier transport [82, 88]. This is depicted in Fig. 12 [35]. In addition to the random timing of the photocarrier generation events, each photocarrier takes a random path across the detector due to, for example, scattering off lattice phonons and impurities. The noise due to this photocarrier scattering can be characterized with an excess noise factor ($F_e$), as is commonly used to describe detectors with internal gain [35], leading to a photocurrent noise variance of $2qI_{avg}\Delta f \times F_e$. Physics-based Monte Carlo (MC) simulations of carrier transport have been utilized to estimate the magnitude of $F_e$ in MUTC photodetectors, and to assess the impact of photocarrier scattering on the microwave phase noise [82]. Within the detector bandwidth, $F_e$ was found to be less than 2, justifying its exclusion in most analyses of unamplified photodetector noise. However, due to the low shot noise contribution to the microwave phase noise in the detection of ultrashort pulses, excess noise due to photocarrier transport will exceed the shot noise and may dominate the phase noise floor.

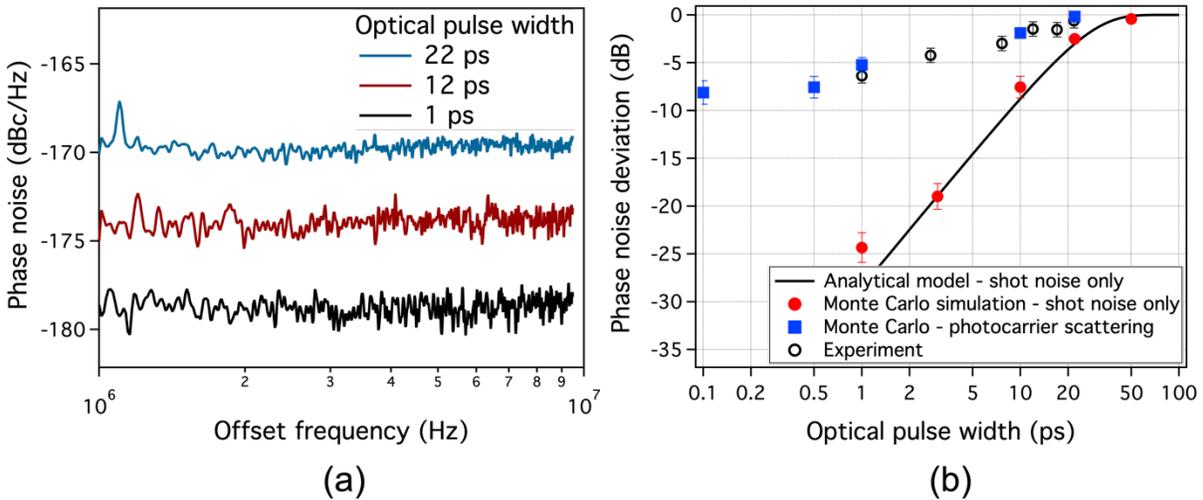

Fig. 14. (a) Measured far-from-carrier phase noise on a 10 GHz carrier derived from photodetection of a train of ultrashort optical pulses. Illumination with shorter optical pulses significantly reduces the white phase noise floor. (b) Comparison of the changes in the white phase noise floor as a function of optical pulse width for different noise models and experimental demonstration. For the shortest optical pulses, modeling indicates the white phase noise floor is dominated by photocarrier scattering.

Figure 14(a) shows measurements of the phase noise floor on a 10 GHz carrier as the optical pulse width illuminating an MUTC photodetector is varied [81]. An average photocurrent in the range of 14 mA to 18 mA was used for all measurements. While there is a clear decrease in the noise as the pulse width is shortened, the noise does not decrease to the level predicted by Eq. 11. Figure 14(b) shows a comparison between Monte Carlo simulations of the photocarrier transport noise and the experimental results, plotted in terms of the phase noise floor deviation from $q/2I_{avg}$, as well as the predicted phase noise due to shot noise only. Monte Carlo simulations and measurement display excellent agreement across more than an order of magnitude variation in optical pulse width, with the shot noise induced phase noise much lower for short optical pulses.

In many cases it is reasonable to assume the statistics on the photocarrier transport is also Poissonian, and can also provide good agreement with experimental results [88]. This simplifies the analysis considerably, since the phase noise due to both shot and photocarrier scattering can be parametrized with the mean electrical pulse width used in place of the optical pulse width. When the optical pulse is short enough that the microwave power will be given by Eq. 1, the phase noise due to combined shot and photocarrier transport noise may be expressed as

$$L_{pt}^{pulse} = \frac{q}{2I_{avg}}\left[1 - \exp\left\{-\left(2\pi f_\mu \tau_e\right)^2\right\}\right] \qquad (15)$$

where we assume Gaussian-shaped electrical pulses with $1/e$ half-width given by $\tau_e$. Importantly, it is the optical pulse convolved with the transit time-limited pulse width that should be used for $\tau_e$, since any purely deterministic filtering of the microwave pulses (due to the external photodetector circuit) will not affect the noise statistics [89]. Of course, the impulse response of a photodetector is typically not Gaussian, but this shape is used here to show the general noise trend. With detailed knowledge of the electrical pulse shape, the formalism of [60] can give more precise predictions of the phase noise.

Though this discussion has identified scattering as a means for photocarrier transport noise, distributed absorption should result in a similar behavior due to the probabilistic nature of the depth within the absorber where a photocarrier is generated. In surface illuminated photodetectors the absorption length is sub-μm and excess phase noise due to distributed absorption is well below that of scattering [82]. However, waveguide detectors can have absorption distributed over several millimeters [90], with the possibility for more significant impact on the phase noise. To our knowledge this "distributed absorption phase noise" has not been experimentally verified.

### 3.1.3 Optical pulse interleavers

When an optical pulse interleaver is used to increase to available microwave power by increasing $f_{rep}$, errors in the interleaver delay impacts the thermal and shot noise floors. As mentioned in Section 2.3, when the microwave frequency of interest is offset from the peak of the interleaver transfer function, the microwave power is reduced relative to its ideal value. An interleaver delay error (Fig. 15(a)) will therefore increase the phase noise floor at the thermal noise limit by an amount equal to the power loss of the microwave harmonic. For example, a single stage interleaver used to generate high power at 10 GHz will produce a power drop of 3 dB from its peak for a delay error of 25 ps. In practice, we find that delay errors of ~1 ps are attainable [65], with negligible impact on the thermal noise limited phase noise. In contrast, the shot noise-limited phase noise floor is much more sensitive to interleaver error. To understand this, we invoke the same time domain picture used above for shot noise. An error in the interleaver delay shifts the pulse time of arrival away from the zero-crossing of the timing reference, leading to higher measured phase noise, as shown in Fig. 15(b). For a single interleaver stage with good power balance, negligible dispersion, and with a delay error $\delta t$, the shot noise-limited phase noise for Gaussian-shaped pulses can be reduced to

$$L^{shot} = \frac{q}{2I_{avg}}\left(2\pi f_\mu\right)^2 \cdot \left[\frac{1}{2}(\delta t)^2 + \left(\tau_{opt}\right)^2\right]. \tag{16}$$

Equation (16) is valid in the limit of ultrashort optical pulses and small $\delta t$, and is derived using Eq. (20) from Ref. [60]. In the presence of photocarrier transport noise, $\tau_{opt}$ can be replaced by $\tau_e$. Thus to avoid a phase noise increase, the delay error should remain below the pulse width. The calculated change in the shot noise-limited phase noise as a function of delay error for a 10 GHz carrier is plotted in Fig. 15(c). The impact of the interleaver delay on the phase noise floor has been verified experimentally [45, 62].

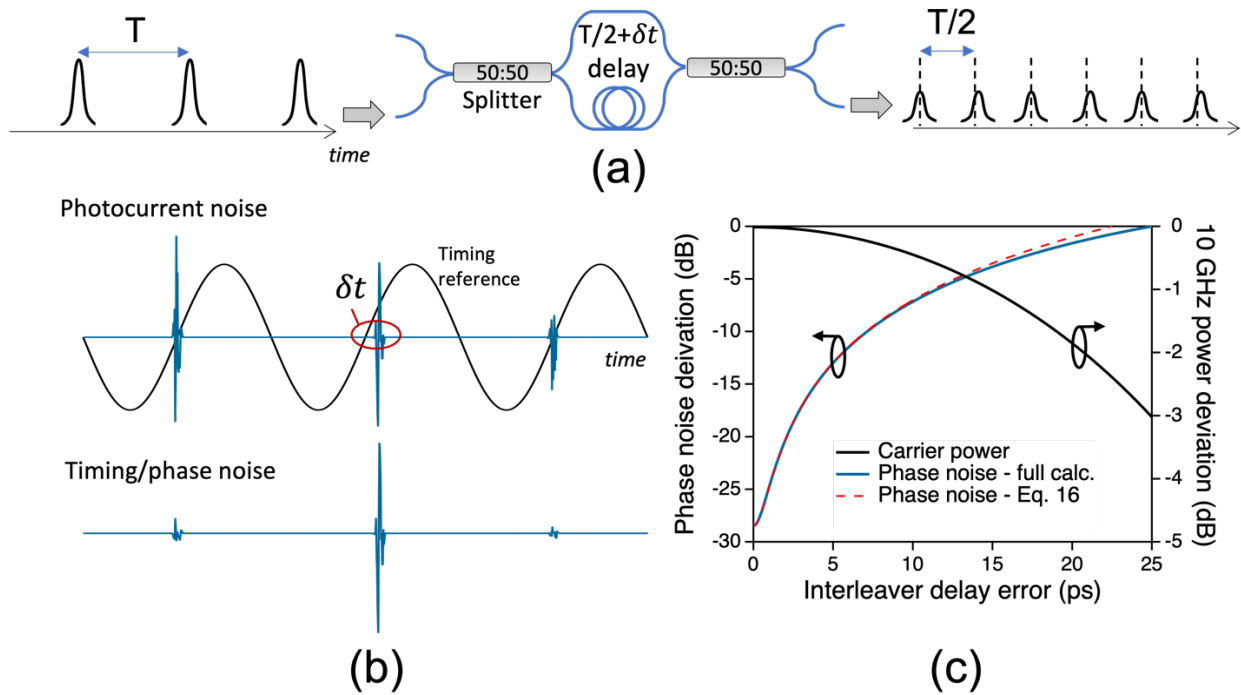

Fig. 15. (a) Schematic of a 1-stage interleaver with delay error $\delta t$. (b) Photocurrent shot noise in the time domain, where the interleaver error produces an offset of the pulse noise with respect to the timing reference. (c) Drop in 10 GHz carrier power (right axis) and increase in shot noise-limited phase noise floor (left axis) as a function of interleaver error. The phase noise deviation calculated from Eq. (16) (dotted red curve) assumes the interleaver error is small, whereas full calculation of the phase noise deviation (solid blue curve) removes this limitation. The assumed pulse width on the detector is 1 ps.

### 3.1.4 Optical amplification

When the microwave phase noise is limited by electronic noise sources, such as thermal noise or photocarrier scattering, optical amplification may be able to improve the overall SNR. However, optical amplification will induce timing jitter over the shot noise limit of the incident pulse train, therefore its impact on the phase noise must be considered. The timing jitter imparted by optical amplification has been examined in long-haul telecommunications systems [91, 92], and this analysis can be adapted for low noise microwave generation. Conceptually, the quantum-limited timing jitter seen at the photodetector may be divided into three parts. First, there will be "direct" timing jitter associated with added spontaneous emission photons that shift the pulse time of arrival. This effect is the short pulse version of signal-spontaneous beat noise as described in [93, 94], where here the beating between the pulse "signal" and spontaneous emission photons occurs only over the duration of the optical pulse. Thus, as with shot noise, the direct jitter from optical amplification scales with the optical pulse width, such that shorter optical pulses will have less direct jitter. Second is Gordon-Haus jitter [95], where spontaneous emission events shift the center of the pulse spectrum that, together with dispersion, cause timing jitter. This is a frequency domain analog to the direct jitter, such that confining the spectrum over a narrower bandwidth should reduce the Gordon-Haus jitter. However, reducing the spectral width to lower the Gordon-Haus jitter can increase the temporal width of the pulses, thereby increasing the direct jitter. Finally, there is a coupling between the direct jitter and the Gordon-Haus jitter for chirped pulses. This coupling term can increase or decrease the total jitter depending on the sign of the chirp and the dispersion of the link between the amplifier and the photodetector.

The problem then is designing the amplifying link between the pulse source and photodetector that balances the direct and Gordon-Haus jitter terms, while also accounting for their coupling in the presence of pulse chirp. The most general cases, where arbitrary pulse shapes, nonlinearity and optical filtering are included, can be solved numerically using the framework described in [91]. Here we look for insight by calculating the timing jitter for a train of unfiltered Gaussian pulses where optical nonlinearity is ignored. Using the linearization approach of [91], the squared timing jitter after a single amplifier followed by a dispersive fiber that includes the direct, Gordon-Haus, and chirp coupling contributions, can be reduced to

$$\sigma_J^2 = \frac{1}{2}\frac{h\nu}{E_{in}}F\tau_{opt}^2 \qquad (17)$$

where $h$ is Planck's constant, $\nu$ is the optical carrier frequency, $E_{in}$ is the pulse energy at the input of the amplifier, and $F$ is an excess noise factor from the optical amplifier, defined as [96]

$$F = 2n_{sp}\left(\frac{G-1}{G}\right). \qquad (18)$$

Here, $n_{sp}$ and $G$ are the population inversion factor and power gain of the amplifier, respectively. Note that $F$ gives the ratio of the quantum-limited timing jitter after the amplifier to the shot noise-limited timing jitter of the input pulses [60, 97]. Thus, despite the complicated interplay between the Gordon-Haus and direct jitter in the presence of pulse chirp, the result is quite simple – the quantum limited timing jitter after amplification is the shot noise limited jitter of the pulses sent into the amplifier multiplied by the amplifier's excess noise factor. To realize the lowest jitter, the dispersion of the link between the optical pulse source and the detector should be designed to attain a transform-limited pulsewidth at the detector. The corresponding microwave phase noise is white, and may be expressed as

$$L_{amp} = \frac{h\nu F}{2P_{in}}\left(2\pi f_\mu \tau_{opt}\right)^2 \qquad (19)$$

where $P_{in}$ is the average optical power at the input of the optical amplifier. By comparing to Eq. (13), we see an increase in phase noise over the shot noise limit by a factor $F$ (for a unity quantum efficiency detector). This noise increase is accepted only when the phase noise is otherwise limited by electronic noise.

Analogous to the shot noise expression in Eq. 13, Eq. 19 gives the phase noise resulting from optical amplification in the short pulse limit. Another approach to the amplifier noise is useful to extend the analysis to longer pulses, for which $f_\mu \tau_{opt} > 0.1$. Here we utilize the shot noise picture of Fig. 13 (d), where now the optical noise power in a 1 Hz bandwidth is given by the spontaneous emission from the amplifier [94] [93], namely

$$S_{ASE} = \frac{1}{2}h\nu \times GF. \qquad (20)$$

The semiclassical interpretation of this equation is a half-photon per mode at the input of the amplifier experiences a gain $G$, with added noise parameterized by $F$. The photocurrent noise power that results from spontaneous emission beating with the optical signal is given by [93, 94]

$$S_{sig-spon} = 2\eta^2 \frac{q^2}{h\nu} G^2 F P_{in} |H(f)|^2 R \tag{21}$$

where $\eta$ is the quantum efficiency of the photodetector. Now we assume this "amplified shot noise" impacts the microwave phase noise in the same manner as has been rigorously derived for shot noise. Under this assumption, the optical amplifier limited phase noise is expressed as

$$L_{amp}^{pulse} = \frac{\eta^2 \frac{q^2}{h\nu} G^2 F P_{in} |H(f)|^2 R}{P_\mu} \left[1 - \exp\left\{-(2\pi f_\mu \tau_{opt})^2\right\}\right]. \tag{22}$$

For short pulses, when the microwave carrier power is given by $P_\mu = 2(GP_{in})^2 \eta^2 (q/h\nu)^2 |H(f)|^2 R$ and $2\pi f_\mu \tau_{opt} \ll 1$, Eq. 22 reduces to Eq. 19, as expected. We compare this to the case of modulated-CW illumination, where the phase noise reduces to

$$L_{amp}^{CW} = 2 \frac{h\nu F}{P_{in}}, \tag{23}$$

again analogous to the shot noise case.

An additional phase noise term ignored in the preceding analysis is due to amplified spontaneous emission (ASE) that directly illuminates the detector. Using Eq. (1) and the photocurrent noise from ASE [93], we estimate the microwave phase noise from this "spontaneous-spontaneous beat noise" in the short pulse limit as

$$L(f) = \frac{1}{8}\left(\frac{h\nu}{P_{in}} F\right)^2 B_{opt}, \tag{24}$$

where $B_{opt}$ is the optical bandwidth of the ASE. This level is typically below other noise terms. More discussion in the relative levels of noise contributions is presented in Section 4.

3.2 Flicker noise

For offset frequencies below ~1 kHz, the photodetector's primary contribution to the microwave phase noise is flicker noise, characterized by a $1/f^a$ power spectrum, where $a \sim 1$. Flicker measurements in high speed photodetectors on a 10 GHz carrier have been reported in the range of -120/f dBc/Hz to -135/f dBc/Hz [34, 98, 99], and phase noise measurements on an optically derived 12 GHz carrier imply a photodiode flicker level near -140/f dBc/Hz [15]. Such low flicker phase noise is on par with the best microwave mixers and amplifiers at these carrier frequencies [100]. In [34], near flicker-limited phase noise was reported for measurement durations exceeding 6000s, with a corresponding fractional frequency stability on a 10 GHz carrier < 6x10[-20] at 1000 seconds of averaging. This level of residual fractional frequency stability can support the best current and next-generation optical clocks [19].

Flicker noise on a photonically generated microwave may be considered as arising from two processes: the baseband flicker noise, generally regarded as arising from conductivity fluctuations due to carrier trapping at surfaces and boundaries [101, 102], and the nonlinear up-conversion of this noise onto the microwave carrier, modulating both its amplitude and its phase. Importantly, flicker noise is only generated while there is a photocurrent. Thus, as with shot noise, the cyclostationary nature of the current under short pulse illumination implies flicker noise may not be equally distributed between amplitude and phase quadratures. This behavior has been examined on a 1 GHz carrier, where the flicker noise was indeed found to reside predominately in the amplitude quadrature for short pulses [103]. Figure 16(a) shows measurements of the photodetector contribution to the amplitude and phase noise of a 1 GHz carrier for long and short pulse widths. For 30 ps-wide current pulses, the separation between amplitude and phase flicker noise is more than 10 dB, whereas the separation goes to nearly zero for 190 ps-wide pulses. The separation between amplitude and phase flicker noise power was calculated under the assumption of cyclostationarity for a range of measured pulse widths and compared to flicker noise measurements. Good agreement between the modeled separation and the noise measurements was obtained, shown in Fig. 16(b). We stress that only the separation between amplitude and phase noise is predicted, not the absolute flicker noise level.

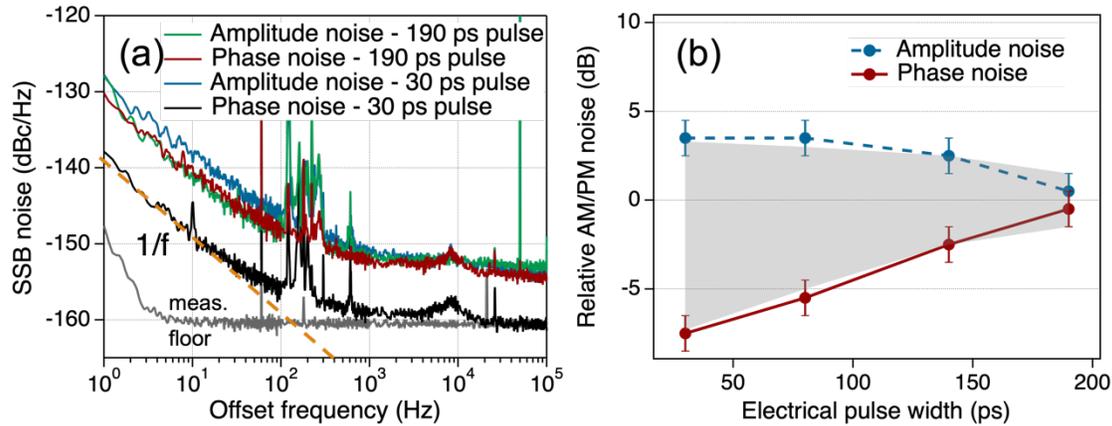

Fig. 16. Flicker noise on a 1 GHz carrier as a function of photodetected electrical pulse width. (a) Single sideband (SSB) noise power spectral densities for 30 ps and 190 ps electrical pulse widths. Noise peaks around 200 Hz are acoustic/seismic in origin. (b) Noise power separation between amplitude and phase in the flicker noise region at 10 Hz offset as a function of electrical pulse width. For the shortest pulses, the flicker noise in the amplitude quadrature is ~10 dB higher than the flicker phase noise. The shaded region shows the expected separation between amplitude and phase noise calculated from the measured electrical pulse widths.

While the observation of cyclostationary flicker noise has brought more insight into the behavior of photodetector flicker, open questions remain in terms of its baseband origins, nonlinear up-conversion, and means of mitigation. In addition to their level imbalance, photodetector amplitude and phase flicker have been shown to be highly correlated [103], implying the noise originates from a common baseband source – an avenue that would be interesting to pursue further. Investigations of the relationship between photodetector saturation and AM-to-PM nonlinearity to flicker may prove fruitful as well. Lastly, we note that details of the device fabrication, such as layer doping levels, have been shown to impact the baseband flicker level of photodetectors [104], though to our knowledge such techniques have not been explored as a means to reduce the flicker noise in high-speed DDR and MUTC detectors.

## 4. Achieving low phase noise: general guidelines and best practices

Given the trade-offs in power, saturation and noise in short pulse detection vs. modulated-CW, and the relative noise of optical amplification compared to photocarrier transport, how does one obtain the best phase noise on a photonically generated microwave? In this section we summarize the results of this paper to create some general guidelines towards obtaining the lowest photodetector phase noise.

*Short pulses can yield lower microwave phase noise than modulated-CW signals, provided the photodiode is not saturated.* Before the onset of saturation, for a given amount of optical power, short pulses yield 6 dB higher microwave power, as well as reduced shot and flicker noise in the phase quadrature. Faster detectors can also be beneficial, since a larger transit-time limited bandwidth will reduce the impact of photocarrier transport noise on the microwave phase. However, the high peak-to-average power ratio of short pulses leads to saturation at lower average photocurrent levels than for modulated-CW signals, diminishing the generated microwave power. In addition to the constraint saturation places on the thermal noise-limited phase noise floor, saturation-induced broadening of the detector's impulse response reduces the shot and flicker noise AM/PM imbalance, increasing the phase noise from these sources. The saturation problem is particularly acute for high harmonics of the pulse repetition rate. In these cases, modulated-CW (or the beat between two optical lines) may be preferable to short pulse illumination.

*The input optical pulse rate should be matched to the desired microwave frequency.* Optical pulse interleavers offer a straightforward path to increase the pulse rate on the photodetector, and can result in orders of magnitude increases in achievable microwave power. Care must be taken to reduce errors in the interleaver delays to avoid increases in the shot noise-limited microwave phase noise. To avoid a phase noise floor increase, the delay errors should be smaller than the transit time-limited electrical pulse width from the detector. In an optically amplified system, the delay error-driven increase in phase noise is magnified by the noise figure of the amplifier, referenced to the optical power at the input of the amplifier.

*Operate at a photocurrent with low AM-to-PM conversion.* The photocurrent of lowest AM-to-PM conversion does not coincide with the highest microwave power. However, when detecting short pulses, matching the repetition rate to the desired microwave frequency, such that $f_\mu$ is not a high harmonic of $f_{rep}$, tends to increase the microwave power that corresponds to an AM-to-PM null. A higher photodetector bias voltage will increase both the microwave saturation power and the AM-to-PM null photocurrent, though one must be careful to avoid thermal damage with an increased bias voltage-photocurrent product. DDR detectors exhibit their first

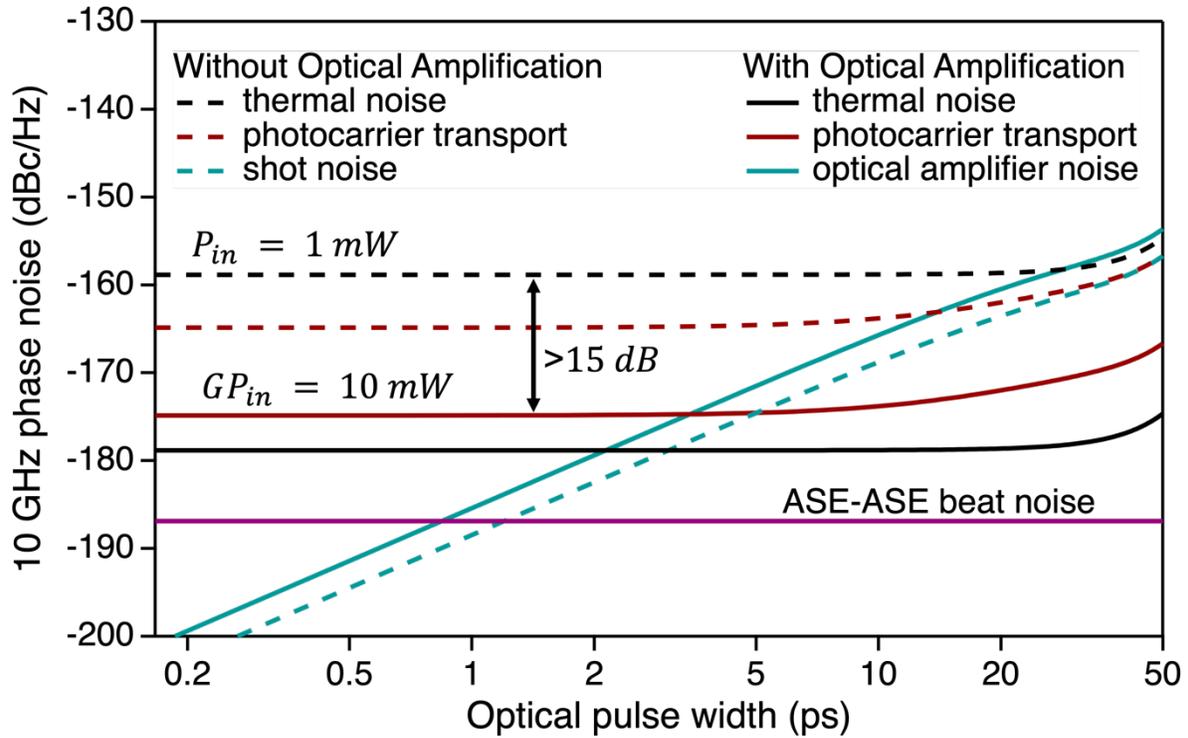

Fig. 17. Summary of the various contributions to the white phase noise floor of a 10 GHz signal derived by photodetection of a train of ultrashort optical pulses. In this example, the optical amplifier's excess noise factor is 5 dB, AM-to-PM conversion is assumed minimal, and the photodetector is unsaturated. Without optical amplification, the phase noise floor is limited by thermal noise. For long pulses, the thermal noise floor increases due to reduced microwave power. With 10 dB amplification, the phase noise floor is limited by photocarrier transport noise for optical pulse widths less than ~ 3 ps, after which the noise is limited by optical amplifier signal-spontaneous beat noise. For an optical pulse width greater than ~25 ps, optical amplification degrades the phase noise performance compared to an unamplified pulse train. For the photocarrier transport noise calculation, the transit time-limited electrical pulse width is 10 ps, and for the ASE-ASE beat noise, the optical bandwidth is 10 THz.

AM-to-PM null at a photocurrent beyond microwave power saturation point where the impulse response of the detector is significantly broadened. For MUTC detectors, the first AM-to-PM null photocurrent is prior to the onset of microwave power saturation and coincides with the shortest electrical pulse width. Thus MUTC detectors can take simultaneous advantage of low AM-to-PM and the short-pulse reduction of flicker and photocarrier transport noise in the microwave phase quadrature.

*Optical amplification can be beneficial when otherwise limited by electronic noise sources.* With short pulse detection, the microwave phase noise floor is typically not shot noise limited. Furthermore, the achievable photocurrent directly from the laser pulse source may not be sufficient to reach a minimum in AM-to-PM conversion. In these cases, optical amplification might be able to reduce the phase noise floor. To compare the phase noise floor with and without optical amplification, we consider changes to the thermal noise floor, changes to the photocarrier transport noise, as well as the added amplifier noise. The ratio of the thermal noise floors is simply given by the ratio of the microwave power with and without optical amplification. For an amplifier with optical power gain $G$, the microwave power ratio is $G^2$ (assuming the microwave power remains unsaturated), leading to a $G^2$ improvement in the thermal noise limited phase noise floor. The larger microwave power with optical amplification also decreases the impact of photocarrier transport noise, though in this case the noise scales inversely with average photocurrent, leading to an improvement of $G$. Lastly, we can compare the photocarrier transport noise floor after optical amplification to the signal-spontaneous beat noise. Under the short pulse approximation, the phase noise ratio is

$$L_{amp}/L_{pt} = \eta FG\ \tau_{opt}^2/\tau_e^2. \tag{25}$$

Thus, for short enough optical pulses and reasonable amplifier gain and noise figure, it is quite possible to still reside in the limit of the photodetector noise despite the extra amplifier noise (again, provided the photodetector is unsaturated).

A fuller picture of the relative contributions of the various noise terms emerges when the reliance on the short pulse approximation is discarded. Figure 17 shows numerical examples for the phase noise of a photonically generated 10 GHz signal as a function of illuminating optical pulse width. The various noise contributions from a shot noise-limited optical pulse train with 1 mW average optical power are shown when the pulse train is directly detected or after the pulse train is optically amplified to 10 mW. In this example, optical amplification improves the phase noise, provided the optical pulse width is below ~25 ps. Without optical amplification, the phase noise floor is dominated by thermal noise, whereas with optical amplification, the noise floor is either dominated by photocarrier transport noise (optical pulse < 3 ps) or by noise from the optical amplifier (optical pulse > 3 ps). For optical pulses ~2 ps and shorter, optical amplification provides a phase noise improvement > 15 dB.

## 5. Conclusion

Taking full advantage of the exquisitely low noise properties of optical sources to produce microwave signals requires exceptionally high-fidelity optical-to-electrical conversion. This paper has reviewed the properties of photodetectors that impact the ability to produce microwave signals with extremely low phase noise. Photodetector saturation, nonlinearities, optical pulse interleavers, and several noise sources have been discussed in detail, with comparisons and trade-offs given between short pulse and modulated-CW illumination. While much of the discussed saturation and nonlinear behavior has focused on DDR and MUTC detectors, the principles apply to any high-speed photodetector. Similarly, the noise analysis is applicable to other photodetector designs as well.

While the sheer number of possibilities in photodetector illumination and response can make it difficult to make broad generalizations, in many cases one can achieve the lowest phase noise microwave signals by detecting trains of optically amplified short pulses. The caveat here is photodetector saturation, exacerbated by high peak powers inherent to short pulses, which reduces the available microwave power and increases the electrical pulse width. Fortunately, saturation can be relieved with optical pulse interleavers, which can be a convenient, low loss method to increase the pulse repetition rate and reduce the peak power on the photodetector. The attainable microwave power can be increased by orders of magnitude in this way, with a similar increase in the microwave power signal-to-noise ratio.

While many aspects of short pulse photodetection are now believed to be understood, there are areas where further study would be quite valuable. Flicker noise, for example, presents an important and particularly challenging problem, since measurements are difficult and the underlying physics is not well understood. Photodetector flicker noise measurements examining carrier frequency, optical power and device fabrication dependencies, as well as links to photodetector nonlinearities such as AM-to-PM conversion, would help advance the understanding of this dominant noise source in the down-conversion of long-term stable optical clock signals. Photodetector operation at cold temperatures could aid in understanding noise properties as well, with the added benefit of supporting experimental endeavors linking photonic systems with superconducting circuits [29, 30, 105]. At room temperature, photocarrier transport noise is believed to be dominated by scattering off lattice photons. At cryogenic temperatures, presumably such scattering is suppressed, and lower noise may result. Baseband flicker noise can also display a temperature dependence [104], and it would be interesting to explore whether microwave flicker phase noise can be improved at cold temperatures.

High fidelity photodetection has helped usher in a new era of extremely phase stable microwave signal generation, with applications in radar, navigation, synchronization, and supporting a new definition of the SI second. The success at microwave frequencies has naturally led to exploration of low noise signal generation at higher frequencies, to date up to 100 GHz [17]. With photodiodes whose bandwidth can reach several hundred GHz [106], and photomixers with THz bandwidth [107], we expect continuing expansion of demonstrations and applications of low noise signal generation based on optical clocks and oscillators.

## Acknowledgements

We are grateful to the many colleagues who have contributed to our understanding and the improvement of high-speed photodetectors over the years, including Haifeng Jiang, Fred Baynes, Eugene Ivanov, Josue Davila-Rodriguez, Dahyeon Lee, Takuma Nakamura, Andreas Beling, Wenlu Sun, J.-D. Deschenes, Jennifer Taylor, Yang Fu, Shubhashish Datta, Abhay Joshi, Thomas Schibli, Craig Nelson, Archita Hati, Antoine Rolland, Jizhao Zang, and Wei Zhang, as well as longtime collaborators in short pulse detection experiments Tara Fortier, Joe Campbell and Scott Diddams. We further thank Bryan Bosworth, and Dahyeon Lee for insightful comments on the manuscript. This work was supported by DARPA, NRL and NIST.